\documentclass[11pt,a4paper,twoside]{article}%
\usepackage[a4paper]{geometry} %
\usepackage{afterpage}

\usepackage{natbib}
\def\citeN{\citet*}
\def\shortciteN{\citet*} 

\newtheorem{theorem}{Theorem}[section]
\newtheorem{lem}{Lemma}[section]
\newtheorem{pro}{Proposition}[section]
\newtheorem{cor}{Corollary}[section]
\newtheorem{conj}{Conjecture}[section]

\newtheorem{rem}{Remark}[section]
\newtheorem{com}{Comments}[section]
\newtheorem{ex}{Example}[section]
\newtheorem{defi}{Definition}[section]
\newtheorem{hyp}{Assumption}[section]

\newcommand{\bt}{\begin{theorem}}\newcommand{\et}{\end{theorem}}
\newcommand{\bl}{\begin{lem}}\newcommand{\el}{\end{lem}}
\newcommand{\bp}{\begin{pro}}\newcommand{\ep}{\end{pro}}
\newcommand{\bcor}{\begin{cor}}\newcommand{\ecor}{\end{cor}}
\newcommand{\bconj}{\begin{conj}}\newcommand{\econj}{\end{conj}}

\newcommand{\bd}{\begin{defi} %
 }\newcommand{\ed}{\end{defi} }
\newcommand{\brem }{\begin{rem}%
}
\newcommand{\erem }{\end{rem}}
\newcommand{\bcom}{\begin{com}%
 }\newcommand{\ecom }{\end{com}}
\newcommand{\brems }{\begin{rem}%
}\newcommand{\erems }{\end{rem}}
\newcommand{\bex}{\begin{ex} %
 }\newcommand{\eex}{\end{ex}}
\newcommand{\bhyp}{\begin{hyp} %
}\newcommand{\ehyp}{\end{hyp}}

\def\proof{\noindent {\it {\textbf{Proof}}}.$\;\,$} 
\newcommand{\dcb}{\begin{array}{lll}}
\newcommand{\dce}{\end{array}}
\newcommand{\ebe}{\begin{enumerate}\setlength{\baselineskip}{13pt}\setlength{\parskip}{5pt}}
\newcommand{\dbe}{\end{enumerate}}

\newcommand{\ibegin}{\begin{itemize}\setlength{\baselineskip}{19pt}\setlength{\parskip}{7pt}}
\newcommand{\iend}{\end{itemize}}

\newcommand{\desb}{\begin{description}}
\newcommand{\dese}{\end{description}}

\usepackage{caption,float}
\usepackage{xspace,cancel}
\usepackage{bold-extra}
\usepackage{bm, enumerate}
\usepackage{dsfont}
\def\href#1#2{#2}
\usepackage[dvips]{graphicx}
\usepackage[dvips,usenames]{color}
\usepackage[xcolor]{pstricks}
\usepackage{latexsym, amsfonts, amssymb, amsmath}
\usepackage{alltt,latexsym, varioref,keyval, nameref, verbatim}

\newtheorem{exo}{Exercice}
\def\bexo{\begin{exo}\rm}
\def\eexo{\end{exo}}
\def\sp{,\,}

\def\G{{\mathbb G}}
\def\Q{{\mathbb Q}}

\def\cB{{\cal {B}}}

\def\cG{{\cal G}}

\newcommand{\beqa}{\begin{eqnarray}}
\newcommand{\eeqa}{\end{eqnarray}}
\newcommand{\beqe}{\beqa\begin{aligned}}
\newcommand{\eeqe}{\end{aligned}\eeqa}
\def\bea{\begin{eqnarray*}}
\def\eea{\end{eqnarray*}}

\def\tilde{\widetilde}
\def\F{{\mathbb F}}\def\F{{\mathfrak F}}
\def\P{{\mathbb P}}

\def\E{{\mathbb E}}

\def\cH{{\cal H}}

\def\gg{\mathbb{G}}

\def\bal{\begin{aligned}}
\def\eal{\end{aligned}}

\newcommand{\COMM}[1]{}
\renewcommand{\COMM}[1]{\begin{quote}\begin{red} \small \sf #1\end{red}\end{quote}}

\def\bal{\begin{aligned}}
\def\eal{\end{aligned}}
\newcommand{\beql}[1]{\beqa\label{#1}\bal}
\newcommand{\eeql}{\eal\eeqa}
\def\bel{\begin{eqnarray*}\begin{aligned}}
\def\eel{\end{aligned}\end{eqnarray*}}

\def\Nu{\tilde{\nu}}\def\Nu{\thisM  }

\def\bal{\begin{aligned}}
\def\eal{\end{aligned}}

\def\beq{\begin{eqnarray}}
\def\eeq{\end{eqnarray}}

\def\E{{\mathbb E}}

\def\cH{{\cal H}}
\def\cS{{\cal S}}

\def\cJ{{\cal J}}

\def\R{\mathbb{R}}

\def\ind{\mathds{1}}

\def\ff{\mathbb{B}}\def\ff{\mathbb{F}}\def\ff{\mathfrak{F}}

\def\tb{\tilde{b}}
\def\thei{i}

\def\bal{\begin{aligned}}
\def\eal{\end{aligned}}

\def\tilde{\widetilde}
\def\hat{\widehat}

\def\ind {1  1}\def\ind{\mathds{1}}
\def\I{\ind}

\def\qqq{\quad\quad\quad}
\def\index#1{}

\newcommand{\finproof}{\rule{4pt}{6pt}}

\def\qr#1{\eqref{#1}}

\def\cG{\mathfrak{G}}
 \def\tb{{\bar{\tau}}}\def\tb{\bar{t}}

\newcommand{\indi}[1]{\I_{\{{#1}\}}}

\def\thisM  t{\tilde{\thisM  }}

\def\FVA{{\rm FVA}\xspace}

\def\KVA{{\rm KVA}\xspace}

\def\VaR{\mathbb{V}\mathrm{a}\mathbb{R}}

\def\loss{\mathcal{L}}\def\loss{L}

\def\FVA{{\rm FVA}\xspace}

\def\cH{{\cal L}}\def\cH{{\cal H}}

\def\KVA{{\hat Y}}\def\KVA{{\rm KVA}\xspace}

\def\cadlag{c\`adl\`ag\xspace}

\usepackage{booktabs}

\def\FVAs{\FVAs }\def\FVAs{\FVA}

\def\eqd{\eq }\def\eqd{=}
\def\eqd{\eq }\def\eqd{=}
\def\eq{:=}\def\eq{=}

\def\ES{\mathbb{ES}}

\newcommand\blu[1]{{\color{blue}#1}}\renewcommand\blu[1]{#1}

\newcommand{\tmop}[1]{\ensuremath{\operatorname{#1}}}

\def\pi{h}
\def\therh{r_h}\def\therh{r}

\def\tp{(t+1)\wedge T}\def\tp{t'}\def\tp{{\overline{t}}}
\def\thisM{M}\def\thisM{\mu}
\def\beta{\delta}

\def\hatY{\hat{Y}}
\def\cY{Z}\def\cY{\mathcal{Y}}

\def\theK{K}\def\theK{\kappa}

\def\then{N}\def\then{n}
\def\them{M}\def\them{m}

\def\thed{D}\def\thed{d} 
\def\thek{l}
\def\thej{j}

\def\cG{\mathfrak{G}}\def\cG{\cF}\def\cG{\F}
\def\Q{\mathbb{Q}}\def\Q{\P}
\def\G{\mathbb{G}}\def\G{\F}
\def\gg{\G}\def\gg{\F}

\def\2l{^l}  \def\2l{_2^l}
   
   \usepackage{amsmath} 
   
   \usepackage[ruled,vlined]{algorithm2e}
\DeclareMathOperator*{\argmin}{arg\,min}

\def\theA{W}\def\theA{A}
\def\theR{\mathbf{R}}\def\theR{\Phi}
\def\theD{D}\def\theD{J}
\def\thed{p}\def\thed{d}
\def\thep{p}
\def\cX{\mathcal{X}}
\def\cX{\mathcal{X}}

\def\Batches{\Batches}\def\Batches{B}
\def\thee{\mu}\def\thee{e}

\newcommand{\Afac}[1][1]{\frac{#1}{1 - \alpha}}
\newcommand{\PolyP}[3][x]{%
    \ifthenelse{\equal{#2}{0}}{P_{#3}(#1)}{%
    \ifthenelse{\equal{#2}{1}}{\tilde{P}_{#3}(#1)}{}}}
\newcommand{\PolyQ}[3][x]{%
    \ifthenelse{\equal{#2}{0}}{Q_{#3}(#1)}{%
    \ifthenelse{\equal{#2}{1}}{\tilde{Q}_{#3}(#1)}{}}}
\newcommand{\Bound}[2]{%
    \ifthenelse{\equal{#1}{0}}{B_{#2}}{%
    \ifthenelse{\equal{#1}{1}}{\tilde{B}_{#2}}{}}}

\newcommand{\EE}[1]{\mathbb{E}^h[|{\rho}^h_{t_{#1}}-\hat{\rho}^h_{t_{#1}}|]}
\newcommand{\EEP}[1]{\mathbb{E}^h[|{Y}^h_{t_{#1}}-\hat{Y}^h_{t_{#1}}|]} 
\newcommand{\ee}[1]{\mathbb{E}^h[\thee_{t_{#1}}]}
\newcommand{\eep}[1]{\mathbb{E}^h[\epsilon_{t_{#1}}]}
\newcommand{\lambdadt}{\Lambda_f \Delta t}
\newcommand{\iidct}{q}
\newcommand{\isumo}{i}
\newcommand{\isumi}{\imath}
\newcommand{\LamPhi}{\Lambda_{\Phi}}\renewcommand{\LamPhi}{{\Lambda^h_{\Phi}}}
\newcommand{\gla}{\left.}
\newcommand{\glb}{\left(}
\newcommand{\glc}{\left[}

\newcommand{\gra}{\right.}
\newcommand{\grb}{\right)}
\newcommand{\grc}{\right]}

\usepackage{hyperref}
\usepackage{placeins}
\hypersetup{
 colorlinks  = true, %
 urlcolor   = black, %
 linkcolor  = black, %
 citecolor  = black %
}
\makeatletter
\renewcommand\@makefnmark{\hbox{\@textsuperscript{\normalfont\color{blue}\@thefnmark}}}
\renewcommand\@makefntext[1]{%
 \parindent 1em\noindent
      \hb@xt@1.8em{%
        \hss\@textsuperscript{\normalfont\@thefnmark}}#1}
\makeatother 

\def\thegamma{\Lambda_f }\def\thegamma{\gamma}

\def\thej{j}
\begin{document}
\title{An Explicit Scheme for Pathwise XVA Computations\footnote{GPU and python implementation  
available   
on https://github.com/BouazzaSE/NeuralXVA.}} 

\author{Lokman Abbas-Turki\textsuperscript{1} \and St\'ephane Cr\'epey\textsuperscript{1} \and Botao Li\textsuperscript{1} \and Bouazza Saadeddine\textsuperscript{1,2,3}}

\date{\today} 
{
{\let\thefootnote\relax\footnotetext{\textsuperscript{1} %
\emph{Laboratoire de Probabilités, Statistique et Modélisation (LPSM), Sorbonne Université et Université Paris Cité, CNRS UMR 8001}}}
{\let\thefootnote\relax\footnotetext{\textsuperscript{2} \emph{LaMME, Univ Evry, CNRS, Universit\'e Paris-Saclay, 91037, Evry, France.}}}
{\let\thefootnote\relax\footnotetext{\textsuperscript{3} \emph{Quantitative Research GMD/GMT Cr\'edit Agricole CIB, 92160 Montrouge.}}} 
 {\let\thefootnote\relax\footnotetext{{{\it Corresponding author:}  {\tt stephane.crepey@lpsm.paris}}}} 
 {\let\thefootnote\relax\footnotetext{{{\it Acknowledgement:} This research has been conducted with the support of the Chair \textit{Capital Markets Tomorrow : Modeling and Computational Issues} 
 under the aegis of the Institut Europlace de Finance,  a joint initiative of Laboratoire de Probabilités, Statistique et Modélisation (LPSM) / Université Paris Cité and  Crédit Agricole CIB.}}}

\maketitle

\begin{abstract} Motivated by the equations of cross valuation adjustments (XVAs) in the realistic case where capital is deemed fungible as a source of funding for variation margin, we introduce a simulation/regression scheme 
for a class of anticipated BSDEs, where the coefficient entails a conditional expected shortfall of the martingale part of the solution. 
The scheme is explicit in time and uses neural network least-squares and quantile regressions for the embedded conditional expectations and expected shortfall computations.
An a posteriori Monte Carlo validation procedure allows assessing the regression error of the scheme at each time step.
The superiority of this scheme with respect to Picard iterations is illustrated in a high-dimensional and hybrid market/default risks XVA use-case.
\end{abstract}

\def\keywordname{{\bfseries Keywords:}}
\def\keywords#1{\par\addvspace\baselineskip\noindent\keywordname\enspace
\ignorespaces#1}

\vspace{2mm}


\section{Introduction}\label{s:intro}

Anticipated BSDES (ABSDEs) are backward stochastic differential equations which coefficient at time $t$
depends on the time-$t$ conditional law of the solution beyond time $t$.  In the seminal ABSDE paper by 
\citeN{PengYang09}, this dependence occurs via a conditional expectation  of the value process $Y$ at some later time. 
Motivated by the equations of cross valuation adjustments (XVAs) in finance,
\citet*{CrepeyElie16} establish the well-posedness of an ABSDE where it occurs via a conditional expected shortfall of a future increment of the martingale part of the solution.
In the present paper we address the numerical solution of such ABSDEs and their XVA application.

The literature on these topics is relatively scarce. In a purely Brownian setup (versus also jumps in our case), \shortciteN{AgarwalDeMarcoGobetLopezNoubiagainZhou18}
 consider an ABSDE involving a conditional expected shortfall as anticipated term
 (by contrast with a conditional expectation in the previous ABSDE literature).
Exploiting the short horizon of the anticipation in the equation (one week in their case versus one year in our XVA application), they devise approximations  by standard 
BSDEs, which allows them to avoid the difficulty posed by the regression of the anticipated terms\footnote{cf.~the beginning of Section 3.2 in \shortciteN{AgarwalDeMarcoGobetLopezNoubiagainZhou18}.}.
The XVA ABSDEs received a first numerical treatment in \citeN{AlbaneseCaenazzoCrepey17b} by nested Monte Carlo\footnote{cf.~also \citeN{AbbasturkiCrepeySaadeddine20}.}, using Picard iterations to decouple the solution from the embedded conditional risk measures and ignoring the conditionings in the latter\footnote{i.e.~computing unconditional risk measures instead of conditional ones.} to avoid multiply nested Monte Carlo.  
The other natural approach to address such problems numerically is regression-based Monte Carlo,
i.e.~iterated regressions 
that are used for cutting the recursively nested levels of Monte Carlo to which a naive implementation of the equations conducts. 
A first take in this direction, still using Picard iterations for decoupling purposes,
was implemented in \citeN{CrepeyHoskinsonSaadeddine2019}, leveraging on the elicitability of the embedded risk measures for learning not only the XVAs, but also conditional value-at-risk for dynamic initial margin calculations and conditional expected shortfall\footnote{that is elicitable jointly with value-at-risk.} for dynamic economic capital calculations. 
Proceeding in this way, \citeN{CrepeyHoskinsonSaadeddine2019} were able to learn the embedded conditional risk measures, instead of treating them numerically as constants in
\citeN{AlbaneseCaenazzoCrepey17b}. 

In this paper we introduce an explicit time-discretization ABSDE scheme which, in conjunction with a refined neural net regression approach for the embedded conditional expectations and risk measures, leads to a direct algorithm for computing pathwise XVA metrics, without Picard iterations. 
The a posteriori error control of Theorem \ref{t:induction} allows the user to overcome the lack of a priori error control inherent to stochastic gradient descent training of neural networks.
A numerical benchmark in a 36 dimensional XVA usecase emphasizes the scalability of the explicit scheme and its superiority with respect to the Picard one.

The paper is outlined as follows.
Section \ref{s:limeqs} recasts the generic ABSDE of \citeN{CrepeyElie16} in a Markovian setup amenable to numerical simulations.
Section \ref{sec:approx} introduces the explicit simulation/regression 
scheme.
Section \ref{s:valid}
deals with its a posteriori   
error control.
Section \ref{sec:the_xva_case} details the XVA
specification of our setup.
Section \ref{ss:numr} provides an XVA numerical benchmark. 
Section \ref{sec:concl} discusses the outputs of the paper in relation with the literature and introduces future research perspectives.

\subsection{Standing Notation}\label{s:ms}

We denote by:
\begin{itemize}
\item 
$|\cdot|$, an Euclidean norm in the dimension of its arguments;
\item  $T\in(0,\infty),$ a constant time horizon;
\item $(\Omega, \mathcal{A},\G,\Q)$, a filtered probability space,  for a 
probability measure $\Q$ on the measurable space $(\Omega, \mathcal{A})$ and  %
a complete and right-continuous filtration $\G=(\cG_t)_{0\le t\le T}$ 
of sub-$\sigma$ fields of  $\mathcal{A}$;
\item $\mathbb{E}$, the $\Q$ expectation,
and $\mathbb{P}_t$, $\mathbb{E}_t$, and  $\mathbb{ES}_t$,
the ($\cG_t,\Q$) conditional probability, expectation, and expected shortfall
at some given quantile level $\alpha\in(\frac{1}{2},(1))$. 
\end{itemize} 
By the latter we mean, for each
$\cG_T$ measurable, 
$\Q$ integrable,
random variable $\ell$,
\beql{iden-ES} 
&\ES _t (\ell)   
 =  \E_t  \big( \ell |\ell \ge q^\alpha_t (\ell)\big) , 
\eeql 
in which $q^\alpha_t(\ell)$ denotes the $(\cG_t,\Q)$
conditional left-quantile\footnote{value-at-risk.} 
of level $\alpha$ of $\ell$.
We recall that\footnote{additionally assuming $\ell$ and $\ell'$ atomless given $\F_t$, without harm for the XVA applications targeted in this work; cf.~e.g.~~\shortciteN[Lemma A.6, Eqn.~(A.16)]{BarreraCrepeyDialloFortGobetStazhynski17}.}, for any $\cG_T$ measurable, $\Q$ integrable random variables $\ell$ and ${\ell'}$,  
 \beql{e:esLip}
 & |\ES  _t (\ell)-\ES  _t  ({\ell'})| 
 \;\leq\; (1-\alpha)^{-1}{{\E}}_t[|\ell-{\ell'}|]. 
 \eeql

\section{Limiting Equations}\label{s:limeqs}

In this section we specify the stochastic differential equations addressed from a numerical viewpoint in later sections.

\subsection{Spaces and Martingale Representation}\label{ss:mrepr}

Given nonnegative integers $\thed$ and $q$,
we denote by $W$, an $(\gg,\Q)$ standard $\thed$ variate 
Brownian motion, 
and $\nu=(\nu^\theK)$, an integer valued random measure 
on $[0,T] \times \{0,1\}^q$\footnote{i.e.~$\nu$ is an $\mathbb{N}\cup\{+\infty\}$ valued 
random measure on $[0,T]\times \{0,1\}^q$ such that $\sum_{\theK\in\{0,1\}^q} \nu^\theK(\{t\} , \omega)\leq 1$ holds $d\mathbb{P}\times dt$ everywhere, cf.
\citet*[Definitions II.1.3, II.1.6.a and II.1.13 on pages 65, 66, and 68]{JacodShiryaev03}.}
with $\Q$ compensatrix
$$d\Nu^\theK_t
=d\nu^\theK_t
-
\thegamma^\theK_t
 dt\sp \theK \in\{0,1\}^q ,
 $$
for some nonnegative
real valued predictable processes $\thegamma^\theK$.
Given any positive integer $\thek$, we introduce:
\begin{itemize}
\item $\cS_2^{\thek},$ the space of $\R^{\thek}$ valued $\ff$ adapted c\`adl\`ag processes $Y$ such that 
\bel
\|Y\|_{\cS_2^{\thek}}^2\eq  \E\big[\underset{0\leq t\leq T}{\sup}{|Y_t|}^2\big]< +\infty;\eel
\item $\cH_2^{\thek},$ the space of $\R^{\thek\otimes \thed}$ valued $\ff$  progressive processes $Z$ such that
$$\|Z\|_{\cH_2^{\thek}}^2\eqd \E\big[ \int_0^T|Z_t|^2 dt \big]< +\infty;$$
\item  
$\tilde{\cH}_2^{\thek},$ the space of $\R^{\thek\otimes  2^q }$ valued $\ff$ predictable processes   
$U$ such that
\begin{align*}
\|U\|_{\tilde{\cH}_2^{\thek}}^2
&\eqd  \E\Big[ \int_0^T  |U |_t^2dt\Big]< +\infty \mbox{ , where }  |U |_t^2= \sum_{\theK
 \in  \{0,1\}^q
} \thegamma_t^\theK  |U _t^\theK |^2.
\end{align*} 
 \end{itemize} 
We use  $%
\int_0^t  U_s
d\Nu_s
$ as  shorthand  for $\sum_{\theK\in\{0,1\}^q}
\int_0^t  U^\theK_s
d\Nu^\theK_s
 $.
\bhyp \label{h:mrp} Every 
$(\F,\P)$  
martingale in $\cS_2^l$
starting from 0
has a representation of the form
 \beql{e:mrp}
\int_0^{\cdot} Z_t dW_t + %
\int_0^{\cdot}  U_t
d\Nu_t
 ,
 \eeql 
for some $Z\in\cH_2^l$ and $U\in\tilde{\cH}_2^l.$
\ehyp
In the case where $\thek=1$ we drop the index $\thek$, e.g.~we write  $\cS_2$ instead of  $\cS_2^1$. 

\subsection{The Markovian Anticipated BSDE} 

In this section, we introduce a Markovian specification of the semimartingale ABSDE of
\citeN{CrepeyElie16}\footnote{see Section \ref{s:intro}.}
 
Given a positive integer $\thep$,
let $X$ in $\cS_2^\thep$ satisfy \beql{SDE}
&dX_t=  b( t, X_t)dt+   \sigma(t, X_t)dW_t 
,
\eeql 
for coefficients $b(t,x)$ and  $\sigma(t, x)$ Lipschitz in $x$ uniformly in $t\in[0,T]$ and with linear growth in $x$. Hence
the SDE \eqref{SDE} is classically well-posed in $S_2^{\thep} $, for any constant initial condition $x\in\R^\thep$.
We write  $\cX=(X,\theD)$, where a $\{0,1\}^q$ valued ``Markov chain like" model component $\theD$\footnote{but with transition probabilities modulated by $X$.} satisfies
 \beql{SDED}
&d\theD_t=  \sum_{\theK\in\{0,1\}^q} (\theK- \theD_{t-})d\nu_t ^\theK 
 \eeql 
 and the compensator $\gamma^\theK_t dt$ of each $d\nu_t ^\theK $ satisfies $\gamma^\theK_t =  \gamma ^\theK (t, \cX_{t-}) $, for some continuous functions $\gamma^\theK  (t, x,k) $ of $(t,x,k)$ such that
  $ \gamma ^\theK (t, x,k) = \indi{k\neq \theK}\gamma ^\theK (t, x,k)$.
Hence
$\nu^\theK_t$ counts the number of transitions of  $\theD$ to the state $\theK$ on $(0,t]$.

We write $f(t, \cX_t,\cdots)$ as shorthand for $f_{\theD_t}(t,X_t,\cdots)$, for any function $f=f_k(t,x,\cdots)$ (e.g.\ $\gamma ^\theK (t, x,k)\equiv \gamma ^\theK_k (t, x)$). 
Given a positive integer $l$, let the terminal condition $\phi=\phi_k (x)$ be for each $k$ an $\R^l$ valued continuous function on $  \R^{\thep} $ , 
the running cost function $f=f_k (t,x, y,\varrho)$  be for each $k$ an $\R^l$ valued continuous function on $[0,T]\times  \R^{\thep} \times \R^l \times \R$, 
 and the conditional expected shortfall
$M\mapsto \ES_{\cdot} (\Phi_{\overline{\cdot} }(M))$ be a 
map  from 
$\cS\2l  $ into the space of $\ff$ predictable\footnote{assuming the raw process $\Phi_{\tp }(M)$ \cadlag in $t$, see \citeN[Lemma 2.1]{CrepeyElie16}.} 
processes,
where\footnote{see e.g.~the last line in \eqref{e:mapping}.}
\beql{e:whirhointr}&
 \Phi_{\tp }(M ):=\theR
  \big(  t;\cX_{[t,\tp ]},  M_{[t,\tp ]}-M_t  \big), 
  \eeql
  for some deterministic maps $t\mapsto \tp  \in[t,T]$\footnote{e.g.~$\tp =(t+1)\wedge T$ in our XVA use case of Sections \ref{sec:the_xva_case}-\ref{ss:numr}.}
  and $\theR=\theR(t,\textbf{x},\textbf{m})$ of time $t$ and \cadlag paths $\textbf{x}$ and $\textbf{m}$ on $[t,\tp ]$ such that $\textbf{m}_t=0$.  
  Assuming $\Phi_{\tp }(M )$ integrable for each time $t$ and $M\in\cS\2l  $,
we consider the following backward stochastic differential equation (BSDE) for $Y$ in $\cS\2l $: 
\beql{BSDE}
& Y_t=\E_t\Big[\phi(\cX_T)+\int_t^{T} f\big(s,\cX_s, Y_s, \ES_s (\Phi_{\overline{s}}( 
 M 
))  
 \big) ds\Big]\sp t\le T, 
\eeql
where $M $, also required to belong to $\cS\2l$, is the canonical Doob-Meyer martingale component of the special semimartingale $Y$.

\begin{hyp}\em\label{hypgeneratorbis}{ \textbf{(i)}}  The function $f=f_k(t,x,y,\varrho)$ is $\Lambda_f $ Lipschitz in $(y,\varrho)$, i.e.
\begin{equation}
    \label{e:fLip}
    \left|  f_k(t,x,y,\varrho) -  f_k(t,x,y',\varrho')\right| \le \Lambda_f \glb\left|y - y'\right|+\left|\varrho-\varrho'\right| \grb; 
\end{equation}
 {%
 \textbf{(ii)}}  The processes
  $     \ES_\cdot| (\Phi_{\overline{\cdot} }(0) |)$
  and  $  | f( \cdot,\cX_\cdot,0,    0)| $
are in $\mathcal{H}_2$;

 {\rm\hfill\break \textbf{(iii)}} 
$\theR$ is $\Lambda_\theR $ Lipschitz with respect to its last argument in the sense that for every $t\in[0,T]$,  
  \beql{e:LipP}|\Phi(t;\mathbf{x},\mathbf{m})-\Phi(t;\mathbf{x},\mathbf{m}')|\le  \Lambda_{\Phi} |\mathbf{m}_{\tp  } - \mathbf{m}'_{\tp  }|   \eeql 
 holds for all \cadlag paths $\mathbf{x},\mathbf{m},  \mathbf{m}'$ on $[t,\tp ]$ such that $\mathbf{m}_{t}=\mathbf{m}'_t=0$.  
 
 \end{hyp}
 
 \brem Assumption \ref{hypgeneratorbis}(iii) strongly points out to the case where $\Phi_{\tp }(M)$ only depends on $M$ through $M_\tp -M_t$, which indeed corresponds to our XVA use case of Sections \ref{sec:the_xva_case}-\ref{ss:numr}.  However, the algorithm of Section \ref{sec:approx} is not restricted to this case and Assumption \ref{hypgeneratorbis}(iii) only yields a sufficient condition for the conclusion of
Proposition \ref{t:absdebis} below to hold.
 \erem

 \bl\label{l:LipP} There exists a positive constant $\Lambda_\rho$ such that
 \beql{Ass_Risk_measure}
 &   | \ES_t (\Phi_{\tp }(M))- \ES_t (\Phi_{\tp }(M'))|^2  \leq  \Lambda_\rho^2 
\E_t \Big[
 \int_{t }^{\tp }  \big( | Z_{s}-Z'_{s}|^2+|U-U'|_{s}^2\big)ds  \Big]   
\eeql
  holds for any $M, M'\in \cS\2l$, where $( Z,U)$ and $( Z',U' )$  in $\cH\2l 
\times \tilde{\cH}\2l $ are the integrands in the martingale representations  \eqref{e:mrp} of $M-M_0$ and $M'-M'_0$. 
\el

\proof By \qr{e:esLip},
we have  
  \bel
 &(1-\alpha) ^{ 2}|\ES_t (\Phi_{\tp }(M))- \ES_t (\Phi_{\tp }(M')) |^2    \leq    \\& 
\left(\E_t 
 \Big[  \theR
  \big(  t;\cX_{[t,\tp]}, M_{[t,\tp]}-M_t  \big)-  \theR
  \big(  t;\cX_{[t,\tp]}, M'_{[t,\tp]}-M'_t  \big)  \Big]\right) ^2 \leq  \\& 
\E_t  \Big[\Big(  \theR
  \big(  t;\cX_{[t,\tp]}, M_{[t,\tp]}-M_t  \big)-  \theR
  \big(t;\cX_{[t,\tp]}, M'_{[t,\tp]}-M'_t  \big)  \Big)^2\Big] 
  ,
\eel 
by the (conditional) Jensen inequality.
Moreover, denoting $\delta M= M-M'$, the Lipschitz condition \eqref{e:LipP} yields 
\bel& 
 \Big(  \theR
  \big(  t;\cX_{[t,\tp]}, M_{[t,\tp]}-M_t  \big)-  \theR
  \big(  t;\cX_{[t,\tp]}, M'_{[t,\tp]}-M'_t  \big)  \Big)^2 \le \Lambda_{\Phi}^2 |\delta M_{\tp } - \delta M_{t}| ^2 ,\eel
  where, with $\delta Z=Z-Z'$ and $\delta U=U-U'$,
  \bel
  &\delta M_{\tp } - \delta M_{t}=\int_t^{\tp } \delta Z_s dW_s +\int_t^{\tp } \delta U_s d\mu_s,
  \eel
  hence 
  \bel
  &|\delta M_{\tp } - \delta M_{t}|^2=\sum_{\iota=1}^l \big(\int_t^{\tp } \delta Z^\iota_s dW_s +\int_t^{\tp } \delta U^\iota_s d\mu_s\big)^2 .
  \eel
Therefore
   \beql{e:Lipfin}
 &(1-\alpha) ^{ 2}|\ES_t (\Phi_{\tp }(M))- \ES_t (\Phi_{\tp }(M')) |^2    \leq     
 \Lambda_{\Phi}^2
 \E_t  \sum_{\iota=1}^l \big(\int_t^{\tp } \delta Z^{\iota}_s dW_s +\int_t^{\tp } \delta U^\iota_s d\mu_s\big)^2 .\eeql 
As a {local} martingale in  $\cS_2$, each process  $\int_t^\cdot \delta Z^\iota_s dW_s +\int_t^\cdot \delta U^\iota_s d\mu_s$ is a square integrable martingale over $[t,{\tp }]$.  
The (conditional) Burkholder inequality\footnote{see e.g. \citet*[Theorem 10.36 and Remark page 285]{HeWangYan92}.} 
applied to this process then 
yields  \bel
 & \E_t  \big(\int_t^{\tp } \delta Z^\iota_s dW_s +\int_t^{\tp } \delta U^\iota_s d\mu_s\big)^2  \le C
\E_t \Big[
 \int_{t }^{\tp }  \big( | \delta Z^\iota_{s} |^2+|\delta U ^\iota|_{s}^2\big)ds  \Big], 
 \eel
so that \eqref{e:Lipfin} entails \eqref{Ass_Risk_measure}.~\finproof\\

\bp\label{t:absdebis} The ABSDE \eqref{BSDE} has a unique special semimartingale solution
$Y$ in $\cS\2l$ with martingale component $M$ in $\cS\2l$. The process $Y$ is the limit in $\cS\2l$ 
of the
Picard iteration defined by $ Y^{0} =0$
and,  
for $\thej\ge 1$,  
\beql{contraction_system-picard}
& Y^{\thej }_t=\E_t\Big[\phi(\cX_T)+\int_t^{T} f\big(s,\cX_s, Y^{ \thej -1 }_s, \ES_s (\Phi_{\overline{s} }(
 M ^{ \thej -1 }
))  
 \big) ds\Big], 
\eeql
where $M ^{ \thej -1 }\in\cS\2l$ is the martingale part of the special semimartingale $Y^{(\thej-1 )}\in\cS\2l$. 
\ep
\proof 
Assumptions \ref{hypgeneratorbis}(i) and   (ii) imply that  
the processes $$ \underset{|y|\leq c}{\sup} |f(\cdot,\cX_\cdot,y, \ES_\cdot \Phi_{\overline{\cdot} }(0))-f( \cdot,\cX_\cdot,0,  \ES_\cdot\Phi_{\overline{\cdot} }(0))|^\frac{1}{2}$$  (for every $c>0$), as well as $ |f( \cdot,\cX_\cdot,0,    \ES_\cdot\Phi_{\overline{\cdot} }(0))|$, are in $\mathcal{H}_2$, which is 
\citep[Assumption 3.2(iii)]{CrepeyElie16}, whereas \citep[Assumption 3.2 (i), (ii), and (iv)]{CrepeyElie16} are implied by  our Assumption \ref{hypgeneratorbis}(i)  and the Lipschitz property of the functions $\phi_k$ combined with the standard a priori bound estimate
$%
 \|X\|_{\cS_2^{\thep}}^2 \le C\left(1+  |x|^2   \right)$ 
on $X$ (with constant initial condition $x$). Moreover,  \eqref{Ass_Risk_measure} corresponds to \citep[Assumption 3.1]{CrepeyElie16}.
Hence \citep[Assumptions 3.1 and 3.2]{CrepeyElie16} hold and the result follows by an application of \citep[Theorem 3.1]{CrepeyElie16}.~\finproof

 \section{The Explicit Simulation/Regression Scheme\label{sec:approx}}
 
 \subsection{Time Discretization}\label{ss:td}

Let there be given a deterministic time-grid $0=t_0< t_1<\cdots <t_n=T$ with mesh size\footnote{maximum time step.} $h$. We write $\Delta t_{i +1  }=t_{i+1}-t_i $. Let $\bar{t}_i$ denote an approximation on the grid of $\overline{t_i}$\footnote{cf.~after \eqref{e:whirhointr}.}.
Let there also be given, on this time grid, simulatable approximations $\cX^\pi$ (assumed Markovian with respect to its own filtration) to $\cX$\footnote{e.g.~the Euler scheme for $X$ and a related approximation for $J$.} and
 $ \theR^h_{\tb_i }   (M^h  )$ to $ \theR_{\tp }   (M  )$, with $M^h$ approximating $M$ on the time-grid and $ \theR^h_{\tb_i }   (M^h  ) $
  of the form\footnote{cf.~\eqref{e:whirhointr}.}
 \beql{e:whirhointrh}& 
 \theR^h \big(  t_i;\cX^h_{\{t_i,\cdots, \tb_i  \}},  M^h_{\{t_i,\cdots, \tb_i  \}}-M^h_{t_i}  \big), 
  \eeql
  for some deterministic map $\theR^h$ of grid times $t_i$ and discrete paths $\textbf{x}^h$ and $\textbf{m}^h$ on $\{t_i,\cdots, \tb_i  \}$ such that $\textbf{m}^h_{ t_i  }=0$.  $\theR^h$ is assumed $\LamPhi$ Lipschitz in the sense mimicking \eqref{e:LipP} that
  \begin{equation}
  \label{e:phihLip}
      \left|\theR^h \big(  t_i;\textbf{x}^h,  \textbf{m}^h \big) - \theR^h \big(  t_i;\textbf{x}^h,  (\textbf{m}^h)' \big)\right| \le\LamPhi\left|\textbf{m}^h_{\bar{t}_i}-(\textbf{m}^h)'_{\bar{t}_i}\right| 
  \end{equation}
holds for any paths $\textbf{m}^h$ and $(\textbf{m}^h)'$ on $\{t_i,\cdots, \tb_i  \}$ such that $\textbf{m}^h_{ t_i  }=(\textbf{m}^h)'_{ t_i  }=0$.

The explicit time discretization for $ (Y , \ES_{\cdot}( \theR_{\overline{\cdot} }   ( M   )))$  (with $M$ the martingale part of the solution  $Y$ to \eqref{BSDE})
is the process
$(Y ^{\pi},\rho^{\pi})$ 
defined at grid times by $Y ^{\pi}_{t_\then }= \phi(\cX^{\pi}_T)%
, \rho ^{\pi}_{t_\then }= \Phi^h_{T}(0)$ and, for $i$ decreasing from $n-1$ to 0,
\beql{approxBSDE} 
&Y ^{\pi}_{t_i}=\E^h_{t_i}\Big[Y ^{\pi}_{t_{i
+1}}+f\big(t_i,\cX^\pi_{t_i},Y ^{\pi}_{t_{i+1}},\rho^{\pi}_{t_{i+1}} \big) \Delta t_{i +1  }\Big]\\
&  \rho^{\pi}_{t_i}=\ES^h_{t_i}\Big(
 \Phi^h_{\tp _i}\big(\underbrace{ Y ^{\pi}_{t_{\cdot}} +\sum_{\imath < \cdot} f(t_\imath , \cX^h_{t_\imath },Y^h_{t_{\imath +1}}, \rho^{\pi}_{t_{\imath +1}}) \Delta t_{\imath +1}}_\text{$M^h$}\big)
  \Big),
\eeql 
where the $\cdot^h$ in $\E^h_{t_i}$ and $\ES^h_{t_i}$ mean that the corresponding conditional expectations and expected shortfalls are in reference to the natural filtration of $\cX^h$.
 
The study of the time-consistency of the scheme, i.e.~the convergence of the $Y^h$ 
to $Y$ as $h$ goes to 0, based on suitable Feynman-Kac representation for the solution of the limiting ABSDE \eqref{BSDE}, 
is left for a separate work (we only provide empirical evidence of the stability of the time  discretization in Section \ref{ss:bench}).
Our main focus in this paper is on the discretization in space of \eqref{approxBSDE}.

\subsection{Fully Discrete Algorithm\label{ss:algos}}

Whenever a process $M^h$ on the time grid is such that
$M^h_{\{t=t_i,\cdots, \tb_i  \}}-M^h_{t_i  }$ is a measurable functional of $( t_i, \cX^h_{t_i }),\ldots, ( \tb_i , \cX^h_{\tb_i }) $ with $\theR^h_{\tb}(M^h)$ square integrable,
 \citeN[Theorem 2.3]{BarreraCrepeyGobetSaadeddine}\footnote{additionally assuming $\theR^h_{\tb}(M^h)$ atomless given $\F_t$.} 
yields, with $\varphi=\varphi_k(t,x)$ and $\psi=\psi_k(t,x)$:
 $\ES^h_t (\theR^h_{\tb}(M^h))=(1-\alpha)^{-1}\psi^*  (t,\cX^h_t)$, where
\beql{e:phi}\psi_{\cdot}^*(t,\cdot)
=\argmin_{\psi_{\cdot}(t,\cdot)\in\cB
}
\E[(\theR^h_{\tb}(M^h) \indi{\theR^h_{\tb }(M^h)\ge \varphi^*(t,\cX^h_t) } - \psi(t,\cX^h_t))^2],\eeql
in which  
\beql{e:varphi}\varphi_{\cdot}^*(t,\cdot)
=\argmin_{\varphi_{\cdot}(t,\cdot)\in\cB
}
\E[( \varphi(t,\cX_t^h)+(1-\alpha)^{-1}  (\theR^h_{\tb}(M^h)-\varphi(t,\cX_t^h))^+],  \eeql 
both
minimizations bearing over the set $\cB 
$ of the Borel functions of $(x,k)$.  

By nonparametric quantile regression estimates
of $ {\varphi}^*  (t,\cX^h_t)$ and $ {\psi}^*  (t,\cX^h_t)$,
we mean any functions $\hat\varphi^*(t,\cdot)$ and $\hat\psi^*(t,\cdot)$ obtained by solving the respective problems \eqref{e:varphi} and \eqref{e:phi} with $\cB
$ approximated by some hypothesis space of functions (e.g. neural networks), $\E$ by the sample mean over a sufficiently large number of independent realizations of $\cX^h$, 
minimization by
 numerical minimization through Adam stochastic gradient descent \citep*{kingma2014adam}, and $\varphi^*$ in  \eqref{e:phi}  by $\hat\varphi^*$.
 
The fully (time and space) discrete counterpart of \eqref{approxBSDE} 
follows by estimating, at each grid time $t=t_i$ going backward, the embedded conditional expectations (resp.~expected shortfalls) through nonparametric least-squares regression against $\cX^h_t$ (resp.~quantile regression against $\cX^h_t$ as explained above). Using the notations $\hat\E^h_{t_i}$ and $\hat\ES^h_{t_i}$ for the embedded conditional expectation and expected shortfall estimators,
we obtain the following ABSDE numerical scheme:  $\hatY ^{\pi}_{t_\then }= \phi(\cX^{\pi}_T)%
, \hat\rho ^{\pi}_{t_\then }= \Phi^h_{T}(0)$ and, for $i$ decreasing from $n-1$ to 0,
\beql{approxBSDEd} 
&\hatY ^{\pi}_{t_i}=\hat\E^h_{t_i}\Big[\hatY ^{\pi}_{t_{i
+1}}+f\big(t_i,\cX^\pi_{t_i},\hatY ^{\pi}_{t_{i+1}},\hat\rho^{\pi}_{t_{i+1}} \big) \Delta t_{i +1  }\Big],\\
&  \hat\rho^{\pi}_{t_i}=\hat\ES^h_{t_i} \Big(
\Phi^h_{\tp _i}\big( 
\underbrace{\hatY ^{\pi} _{t_\cdot} +\sum_{\imath  < \cdot} f(t_\imath , \cX^h_{t_\imath },\hat{Y}^{\pi }_{t_{\imath+1} }, \hat{\rho}^{\pi }_{t_{\imath+1} } )\Delta t_{\imath +1}
 }_\text{$\hat{M}^{h}$}
\big) 
 \Big).
\eeql
Note that
$$\hat{M}^{h}_{t_{i}}=\hat{M}^{h}_{t_{i+1}}
+ \hatY ^{\pi}_{t_{i}}- \cY ^{\pi}_{t_{i+1}}\mbox{ ,  where }
\cY ^{\pi}_{t_{i+1}}=\hatY ^{\pi}_{t_{i+1}}  +f\big(t_i,\cX^\pi_{t_i},\hatY ^{\pi}_{t_{i+1}},\hat\rho^{\pi}_{t_{i+1}} \big) \Delta t_{i +1  ,}$$
which is exploited line 19 of Algorithm \ref{alg:explicit_backward_algo}
to compute $\hat{M}^{h}_{t_i}$ iteratively (for decreasing $i$).

As in \citeN{AbbasturkiCrepeySaadeddine20}, given weight matrices $\theA^{[L+1]}\in\mathbb{R}^{1\times u},\dots,\theA^{[\ell]}\in\mathbb{R}^{u\times u},\dots,\theA ^{[1]}\in\mathbb{R}^{u\times  (\thep+ q) }$, biases $b^{[L+1]}\in\mathbb{R},\dots,b^{[\ell]}\in\mathbb{R}^u,\dots,b^{[1]}\in\mathbb{R}^u$, and a scalar non-linearity $\varsigma$ applied element-wise, let, for every $z=(x,k)\in\mathbb{R}^{ \thep }\times\mathbb{R}^{ q}\equiv\mathbb{R}^{ \thep + q}$, 
\[
\begin{aligned}
&\zeta^{[0]}(z;\theA ,b) = z\\
&\zeta^{[\ell ]}(z;\theA ,b)= \varsigma(\theA ^{[\ell ]}\zeta^{[\ell-1]}(z;\theA ,b)+b^{[\ell]}),\quad \ell=1, \dots, L \\
& \zeta^{[L +1]}(z;\theA ,b) = \theA ^{[L +1]}\zeta^{(L )}(z;\theA ,b)+b^{[L +1]},
\end{aligned}
\]
where $\theA $ and $b$ denote the respective concatenations of the $\theA ^{[\ell]}$ and of the $b^{[\ell]}$. The function $\mathbb{R}^{\thep+ q}\ni z\mapsto\zeta^{[L +1]}(z;\theA ,b)\in\mathbb{R}$ then implements a neural network with $L$ hidden layers, $u$ neurons per hidden layer, $\varsigma$ as an activation function applied on each hidden unit, and no non-linearity at the output layer. Using the corresponding parameterized set of functions  $\mathbb{R}^{\thep+ q}\ni z\mapsto\zeta^{[L +1]}(z;\theA ,b)\in\mathbb{R}$ as hypothesis space in the description following \eqref{e:phi}-\eqref{e:varphi},
we obtain
Algorithm \ref{alg:explicit_backward_algo},
 using  Algorithm \ref{alg:baseline_algo} as an elementary learning block therein. At the beginning of the algorithms, i.e.~before any learning is performed, the weights are initialized randomly the way explained in \citeN{GoodfellowBengioCourville2017}. In
our numerical experiments, we use softplus activation functions in the hidden layers, in combination with the related weight initialization scheme in \citeN{he2015delving}.

{
\setcounter{algocf}{-1}
\begin{algorithm}[h]
\small
\LinesNumbered
\SetAlgoLined
\SetKwInOut{AlgName}{name}
\SetKwInOut{Input}{input}\SetKwInOut{Output}{output}
\AlgName{NNRegress}
\Input{$\{(\chi^{\jmath}, \xi^{\jmath}),  \jmath\in\cJ \}$, a partition $\Batches$ of $\cJ $, a number of epochs $E\in\mathbb{N}^{\star}$, a learning rate $\eta>0$, initial values for the network parameters $\theA $ and $b$, type of regression \textit{regr}}
\Output{Trained parameters $\theA $ and $b$} 
define $\displaystyle\mathcal{L}(\theA , b, \text{batch})= \left\{\begin{aligned}&\frac{1}{|\text{batch}|}\sum_{\jmath\in\text{batch}} (\zeta^{[L+1]}(\chi^{\jmath}; \theA , b)-\xi^{\jmath})^2 &\text{if }\textit{regr}=\text{ls}\\ &\frac{1}{|\text{batch}|}\sum_{\jmath\in\text{batch}} (\xi^{\jmath}-\zeta^{[L+1]}(\chi^{\jmath}; \theA , b))^++(1-\alpha)\zeta^{[L+1]}(\chi^{\jmath}; \theA , b) &\text{if }\textit{regr}=\text{qle}\end{aligned}\right.$\\
 \For(\tcp*[f]{loop over epochs}){$\text{epoch} = 1, \dots, E$}{
  \For(\tcp*[f]{loop over batches}){$\text{batch} \in \Batches$}{
   $\theA, b \leftarrow \theA - \eta \nabla_{\theA} \mathcal{L}(\theA, b, \text{batch}),  b - \eta \nabla_{b} \mathcal{L}(\theA, b, \text{batch})$\\
  }
 }
\caption{Elementary learning block for least squares (\text{ls}) or quantile (\text{qle}) neural net regressions of cash flows
$\xi^\jmath$ against features $\chi^\jmath$ indexed by $\jmath\in\cJ$}\label{alg:baseline_algo}
\end{algorithm}

\begin{algorithm}[!htbp]
\small
\LinesNumbered
\SetAlgoLined
\SetKwInOut{AlgName}{name}
\SetKwInOut{Input}{input}\SetKwInOut{Output}{output}
\Input{{current state $ \mathcal{X}_{0}=x$}$, \{\{\mathcal{X}_{t_i}^{h, \jmath},(1) \leq i \leq n\},
\phi(\cX^{h, \jmath}_T), 
 \jmath\in\cJ \}$, a partition $\Batches$ of $\cJ $, a number of epochs $E\in\mathbb{N}^{\star}$, 
 a learning rate $\eta>0$} 
\Output{{$\widehat{Y}^h_0$ and learned parameters $\{(\theA_{i}^{\VaR}, b_{i}^{\VaR}), (\theA_{i}^{\ES}, b_{i}^{\ES}), (\theA _{i}^{\iota}, b_{i}^{\iota}),\iota \in\{1, \dots, l\}, i \in\{1, \dots, n\}\}$}} 
For all $\jmath\in\cJ $, let $y^{\jmath}\in\mathbb{R}^{l}$ and, for each $i=0\cdots n,$ $\hat{M}^{h,\jmath}_{t_i}=0\in\mathbb{R}^{l}$

Initialize parameters
 $\{(\theA_{n}^{\VaR}, b_{n}^{\VaR}), (\theA_{n}^{\ES}, b_{n}^{\ES})\}$ of the networks approximating the $\VaR$ and $\ES$ at terminal time-step $n$

Initialize the parameters
 $\{(\theA _{n}^{\iota}, b_{n}^{\iota}),\iota \in\{1, \dots, l\}\}$ of the least-squares networks, indexed by $\iota\in\{1, \dots, l\}$, at terminal time-step $n$

\lForEach*{$\jmath\in\mathcal{J}$}{
 $ y^{\jmath} \leftarrow
 \phi(\cX_T^{h, \jmath})$,
 $\hat{M}^{h,  \jmath}_{t_n} \leftarrow 0$ 
}

\For{$i=n-1 \dots 1$}{
\lForEach*{$\jmath\in\cJ $}{$\xi^{\jmath} \leftarrow 
 \theR^h_{{\bar t}_i}  (  \hat{M}^{h,\jmath} )$}
 \tcp{We first learn the $\VaR$ using a quantile regression}

 ${\theA _{i}^{ \VaR}, b_{i}^{ \VaR} \leftarrow \text{NNRegress}(\{(\mathcal{X}_{t_i}^{h, \jmath}, \xi^{\jmath}), \jmath\in\cJ \}, \Batches, E, \eta, \theA _{i+1}^{ \VaR}, b_{i+1}^{ \VaR}, \text{qle})}$\tcp{Then we deduce the $\ES$ using an $\text{ls}$ regression}

 $\xi^{\jmath} \leftarrow \frac{1}{1-\alpha}\xi^{\jmath} \mathbf{1}_{\{\xi^{\jmath} \geq \zeta^{[L+1]}(\mathcal{X}_{t_i}^{h, \jmath}; \theA^{ \VaR}_{i}, b^{ \VaR}_{i}) \}}$
 
$\theA _{i}^{ \ES}, b_{i}^{ \ES} \leftarrow \text{NNRegress}({\{(\mathcal{X}_{t_i}^{h, \jmath}, \xi^{\jmath}), \jmath\in\cJ \}, \Batches, E, \eta, \theA _{i+1}^{ \ES}, b_{i+1}^{ \ES}, \text{ls}})$

\tcp{We compute the integrand $\xi$ that needs to be projected to get the solution of the BSDE at the current time step}

$\varrho^{\jmath}\leftarrow\zeta^{[L+1]}(\mathcal{X}_{t_{i+1}}^{h, \jmath}; \theA _{i+1}^{ \ES}, b_{i+1}^{ \ES})$

$\xi^{\jmath} \leftarrow  y^{\jmath}+f(t_i, \mathcal{X}_{t_i}^{h, \jmath}, y^{\jmath}, \varrho^{\jmath}) \Delta t_{i+1}$

\For{$\iota=1 \dots l$}{
    ${\theA _{i}^{\iota}, b_{i}^{\iota} \leftarrow \text{NNRegress}(\{(\mathcal{X}_{t_i}^{h, \jmath}, \xi^{\jmath}_{\iota}), \jmath\in\cJ \}, \Batches, E, \eta, \theA _{i+1}^{\iota}, b_{i+1}^{\iota}, \text{ls})}$
  }

\tcp{Compute $\hat{M}^h_{t_i}$}

\For{$\jmath\in\mathcal{J}$}{
\For{$\iota=1 \dots l$}{
 $ y_{\iota}^{\jmath} \leftarrow \zeta^{[L+1]}(\mathcal{X}_{t_i}^{h, \jmath}; \theA _{i}^{\iota}, b_{i}^{\iota})$
}

$\hat{M}^{h,\jmath}_{t_i} \leftarrow \hat{M}^{h,\jmath}_{t_{i+1}}+y^{\jmath}-\xi^{\jmath}$

}
}

    \For{$\jmath\in\mathcal{J}$}{
$\varrho^{\jmath}\leftarrow\zeta^{[L+1]}(\mathcal{X}_{t_{1}}^{h, \jmath}; \theA _{1}^{ \ES}, b_{1}^{ \ES})$

$\xi^{\jmath} \leftarrow  y^{\jmath}+f( 0, x, y^{\jmath}, \varrho^{\jmath}) \Delta t_{1}$}
$\widehat{Y}^h_0 \leftarrow sampleMean(\xi)$

\caption{Explicit backward learning scheme for the ABSDE \eqref{BSDE} based on time-discretized simulated paths $\cX^{h,\jmath}$ of $\cX$ indexed by $\jmath\in\cJ$}
\label{alg:explicit_backward_algo}
\end{algorithm}

\section{A Posteriori Analysis of the Regression Error \label{s:valid}}

In the case of dynamic programming based simulation/regression BSDE schemes such as \eqref{approxBSDEd} ,
the usual spatial error analysis strategy consists in controlling the accumulation, over (discrete) time $i$ decreasing from $n-1$ to 0, of three error components
\citep[Eqn.~(VIII.3.8)]{gobet2016monte}:
(i) a \textit{bias} between the function $u^i$ representing $Y^h_{t_i}$ as $u^i(\cX^h_{t_i})$ (for a suitable measurable function $u^i=u^i_k(x)$) and the hypothesis space of functions in which $\hat{Y}^h_{t_i}$ is sought after, (ii) a \textit{``variance"} in the sense of a regression estimation error, and (iii) a  term of \textit{propagation} at time $i$ of the error at time $i+1$. 
This is at least the strategy in the standard case where the embedded  conditional expectations are estimated by parametric
least-squares regressions that can be performed exactly, for instance by singular value decomposition \citep[Section VIII.2.2]{gobet2016monte}.
Neural net parameterizations for the target functions (conditional expectations but also expected shortfalls in the case of our ABSDEs) instead lead to ``nonlinear regressions" that can only be performed by numerical, nonconvex minimization. When state-of-the-art, fine-tuned,  Adam  variants
of stochastic gradient descents are used, the ensuing minimization can be
quite efficient numerically. But
there is no known learning algorithm solving such nonconvex minimization problems with an a priori error bound. Hence, when the learning iteration terminates, we do not have any guarantee on the quality of the approximation. 
In other words, 
there is a fourth \textit{numerical minimization} error component on top of the three other ones in the above and  this fourth error component cannot be controlled ex ante.

All in one, no exhaustive a priori error control can be hoped for on such problems. 
However, we can still assess the regression error of the schemes by an a posteriori 
twin Monte Carlo validation procedure, which is the topic of this section.

\subsection{A Posteriori Validation of the Local Regression Error\label{ss:valid}}

\newcommand{\PhiLast}{\Phi^*}\renewcommand{\PhiLast}{\Phi}
For notational simplicity
we assume a uniform time step $\Delta t_{i +1  }=\Delta t$, in the  XVA motivated case
(see above \eqref{e:EC}) $\tp =(t+1)\wedge T, \tb_i =t_{(i+\them )\wedge n},$ with  
$\them=  [\frac{1}{\Delta t}]$. Hence we have in \eqref{approxBSDEd}
 (cf. \eqref{e:whirhointrh}):
\bel&\Phi^h_{\tb_i 
}\big( \hatY ^{\pi} _{t_\cdot} +\sum_{\imath  < \cdot} f(t_\imath , \cX^h_{t_\imath },\hat{Y}^{\pi }_{t_{\imath+1} }, \hat{\rho}^{\pi }_{t_{\imath+1} } )\Delta t 
\big)=\\&\qqq\theR^h \Big(  t_i;\cX^h_{\{t_i,\cdots, \tb_i  \}}, \hatY ^{\pi}_{ t_\cdot}-\hatY ^{\pi}_{ t_i   }  +\sum_{ i\le \imath  < \cdot} f(t_\imath , \cX^h_{t_\imath },\hat{Y}^{\pi }_{t_{\imath+1} }, \hat{\rho}^{\pi }_{t_{\imath+1} })\Delta t  \sp i \le \cdot \le (i+m)\wedge n \Big). \eel  
On top of \eqref{approxBSDEd}, 
we also define the auxiliary %
scheme $\tilde{Y} ^{\pi}_{t_\then }  =  \phi(\cX^{\pi}_T), \tilde{\rho}_{t_{n}}^h=0$ and, for $i$ decreasing from $n-1$ to 0,
\beql{e:Yrt}
\tilde{Y}^{\pi}_{t_i} &= \mathbb{E}^h_{t_i}[\hat{Y}^{\pi}_{t_{i+1}}+f(t_i, \cX^{\pi}_{t_i}, \hat{Y}^{\pi}_{t_{i+1}}, \hat{\rho}_{t_{i+1}}^h)\Delta t]\\
\tilde{\rho}_{t_{i}}^h &= \mathbb{ES}^h_{t_{i}}[\Phi^h_{\tb_i 
}\big( \hatY ^{\pi} _{t_\cdot} +\sum_{\imath  < \cdot} f(t_\imath , \cX^h_{t_\imath },\hat{Y}^{\pi }_{t_{\imath+1} }, \hat{\rho}^{\pi }_{t_{\imath+1} } )\Delta t ].
\eeql
Let \beql{e:epsmu}\epsilon_{t_i} =|\tilde{Y}^h_{t_i}-\hat{Y}^h_{t_i}|\sp
e_{t_i}  =|\tilde{\rho}^h_{t_i}-\hat{\rho}^h_{t_i}|
.\eeql
Proceeding as in \citeN[Section 5.1]{AbbasturkiCrepeySaadeddine20}, one can estimate $\mathbb{E}^h[\epsilon_{t_i}^2]$  {\itshape without computing $\tilde{Y}^{\pi}_{t_i}$}, by Monte Carlo using a so called twin simulation scheme. The latter consists in a Monte Carlo estimation of $\mathbb{E}^h[\epsilon_{t_i}^2]$ based on the following representation:
\bl\label{lem:twin}
Let there be given two copies $\hat{Y}^{\pi,(1)}_{t_{i+1}}$ and $\hat{Y}^{\pi,(2)}_{t_{i+1}}$ of $\hat{Y}^{\pi}_{t_{i+1}}=u
^{i+1}
(\cX^h_{t_{i+1}})$ and
$\hat{\rho}^{\pi,(1)}
_{t_{i+1}}$ and $\hat{\rho}^{\pi,(2)}
_{t_{i+1}}$ of $\hat{\rho}
^{\pi}_{t_{i+1}}=v
^{i+1}(\cX^h_{t_{i+1}})$, where $ u^{i+1}$ and $ v^{i+1}$ are the regressed functional forms of $\hat{Y}^{\pi}_{t_{i+1}}$ and $\hat{\rho}^{\pi}_{t_{i+1}}$, the two copies 
$\hat{Y}^{\pi,(1)}_{t_{i+1}}$ and $\hat{Y}^{\pi,(2)}_{t_{i+1}}$ of $\hat{Y}^{\pi}_{t_{i+1}}$
are
independent conditionally on $\cX^{\pi}_{t_i}$\footnote{i.e.~one simulates two independent realizations of $\cX^h_{t_{i+1}}$, given the same starting point $\cX^h_{t_i}$, and then takes their images by the learned functionals $u
^{i+1}$ and $v
^{i+1}$.}, and so are $\hat{\rho}^{\pi,(1)}
_{t_{i+1}}$ and $\hat{\rho}^{\pi,(2)}
_{t_{i+1}}$.
Then
\beql{e:twins}\mathbb{E}^h& [\epsilon_{t_i}^2] = \mathbb{E}^h\left[|\hat{Y}^{\pi}_{t_{i}}|^2\right.\\ &\left.-(\hat{Y}^{\pi}_{t_{i}})^{\top}\left(\hat{Y}^{\pi,(1)}_{t_{i+1}}+f(t_i, \cX^{\pi}_{t_i}, \hat{Y}^{\pi,(1)}_{t_{i+1}},\hat{\rho}_{t_i+1}^{h,(1)})\Delta {t}\right.+ \hat{Y}^{\pi,(2)}_{t_{i+1}}+f(t_i, \cX^{\pi}_{t_i}, \hat{Y}^{\pi,(2)}_{t_{i+1}},\hat{\rho}_{t_i+1}^{h,(2)})\Delta {t}\right)\\ &\left.+\glb\hat{Y}^{\pi,(1)}_{t_{i+1}}+f(t_i, \cX^{\pi}_{t_i}, \hat{Y}^{\pi,(1)}_{t_{i+1}},\hat{\rho}_{t_i+1}^{\pi,(1)})\Delta {t})^{\top}(\hat{Y}^{\pi,(2)}_{t_{i+1}}+f(t_i, \cX^{\pi}_{t_i}, \hat{Y}^{\pi,(2)}_{t_{i+1}},\hat{\rho}_{t_i+1}^{\pi,(2)})\Delta {t}\grb\right].\eeql
\el
\proof This follows by application of the following identity, which holds by application of the tower rule for any
Borel function $\varphi:\mathbb{R}^p\times\{0,1\}^q\rightarrow\mathbb{R}$ such that $\varphi(\cX^{\pi}_{t_i})$ is square integrable (e.g.~any component of the vector  $\hat{Y}^{\pi}_{t_{i}}$)
and any
square integrable random variable $\xi$ (e.g.~any component of the vector  $\hat{Y}^{\pi }_{t_{i+1}}+f(t_i, \cX^{\pi}_{t_i}, \hat{Y}^{\pi }_{t_{i+1}},\hat{\rho}_{t_i+1}^{h })\Delta {t}$) with
copies $\xi^{(1)}$ and $\xi^{(2)}$ independent conditionally on $\cX^{\pi}_{t_i}$\footnote{the conditional independence means that for any Borel bounded functions $\phi_1$ and $\phi_2$, we have $\mathbb{E}[\phi_1(\xi^{(1)})\phi_2(\xi^{(2)})|\cX^{\pi}_{t_i}] = \mathbb{E}[\phi_1(\xi^{(1)})|\cX^{\pi}_{t_i}]\mathbb{E}[\phi_2(\xi^{(2)})|\cX^{\pi}_{t_i}]$.}:
\beql{e:val1}\mathbb{E}[(\varphi(\cX^{\pi}_{t_i})-\mathbb{E}[\xi|\cX^{\pi}_{t_i}])^2] = \mathbb{E}[\varphi(\cX^{\pi}_{t_i})^2-\varphi(\cX^{\pi}_{t_i})(\xi^{(1)}+\xi^{(2)})+\xi^{(1)}\xi^{(2)}].~\finproof\eeql

\noindent
Proceeding in this way, we obtain an a posteriori Monte Carlo procedure to assess
the local regression error $\sqrt{\mathbb{E}^h[ \epsilon^{2}_{t_i} ]}$ 
at each pricing time step\footnote{see e.g.~Figures \ref{fig:local_l2_err_via_twin_fva}-\ref{fig:local_l2_err_via_twin_kva} in Section \ref{ss:numr}.}.
An alternative twin Monte Carlo simulation procedure is also available from \citet*[Section 4.4]{BarreraCrepeyGobetSaadeddine} to estimate an upper bound on $\mathbb{E}^h[e^2_{t_i}]$ (cf.~\eqref{e:epsmu})  without having to compute the 
$\tilde{Y}^h_{t_i}$ and $\tilde{\rho}^h_{t_i}$. But in this expected shortfall case (as opposed to conditional expectation regarding $\mathbb{E}^h[ \epsilon^{2}_{t_i} ]$), this procedure only yields an upper bound, and one which also entails a lower bound $c$ on the density of the law of the labels given the features of the learning problem  (see \citet*[Eqn.~(4.7)]{BarreraCrepeyGobetSaadeddine}), i.e. of $\Phi^h_{\tb_i 
}\big( \hatY ^{\pi} _{t_\cdot} +\sum_{\imath  < \cdot} f(t_\imath , \cX^h_{t_\imath },\hat{Y}^{\pi }_{t_{\imath+1} }, \hat{\rho}^{\pi }_{t_{\imath+1} } )\Delta t$
given $\cX^h_{t_i}$ in our $e_{t_i}$ case, for which we do not know $c$. All in one, the main practical use of the above is for assessing
the local regression error $\sqrt{\mathbb{E}^h[ \epsilon^{2}_{t_i} ]}$  
at each pricing time step. But when considered in conjunction with the corresponding $L^2$ training losses which could also be used as a naive (overconservative) error estimate (cf.~Figures \ref{fig:local_l2_err_via_twin_fva}-\ref{fig:local_l2_err_via_twin_kva} in Section \ref{ss:numr}), this is already very useful to guide the design of the learning procedure.
In the case where the error  $\sqrt{\mathbb{E}^h[ \epsilon^{2}_{t_i} ]}$  is not good enough (or too close to the $L^2$ training loss), one can improve the stochastic gradient descent, in first attempt, and then act on the hypothesis spaces, e.g., in the case of neural networks, try to train with more layers/units or better architectures.

\subsection{Accumulation in Time of the Regression Errors}

In the standard BSDE case where $f_k(t,x,y,\varrho)=f_k(t,x,y)$, i.e.~when there is no dependence of the coefficient of the BSDE on $\rho$, we have
by $\Lambda_f $ Lipschitz continuity of $f_{k} (t, x, y )$ with respect to $y$:
\[\mathbb{E}^h[|Y_{t_i}^h-\tilde{Y}_{t_i}^h|] \leq (1+\Lambda_f  \Delta t)\mathbb{E}^h[|Y^h_{t_{i+1}}-\hat{Y}^h_{t_{i+1}}|] \]
and the triangular inequality yields
\beql{e:err_bound_Y}
\mathbb{E}^h[|Y_{t_i}^h-\hat{Y}_{t_i}^h|] \leq (1+\Lambda_f  \Delta t)\mathbb{E}^h[|Y^h_{t_{i+1}}-\hat{Y}^h_{t_{i+1}}|]+\mathbb{E}^h[|\hat{Y}_{t_i}^h-\tilde{Y}_{t_i}^h|] .
\eeql
Hence, using the inequality $\mathbb{E}^h[\epsilon_{t_i}] \leq \sqrt{\mathbb{E}^h[\epsilon_{t_i}^2]}$:
\beql{e:major}
\mathbb{E}^h[|Y ^{\pi}_{t_i}-\hat{Y}^{\pi}_{t_i}|] \leq \sum_{\imath=i}^{\then -1} (1+\Lambda_f \Delta {t})^{\imath-i}\sqrt{\mathbb{E}^h[ 
\epsilon_{t_\imath}^2]},
\eeql 
where each $\mathbb{E}^h[\epsilon_{t_\imath}^2] $ in \eqref{e:major} can be computed by twin Monte Carlo
based on \eqref{e:twins}\footnote{ignoring the $\hat{\rho}_{t_i+1}^h$ there.}, for
two copies $\hat{Y}^{\pi,(1)}_{t_{i+1}}$ and $\hat{Y}^{\pi,(2)}_{t_{i+1}}$ of $\hat{Y}^{\pi}_{t_{i+1}}$
 independent conditionally on $\cX^{\pi}_{t_i}$. 

In the anticipated case where $f$ also depends on $\varrho$, the analogous propagation of the local regression error terms $\epsilon_{t_\imath}$ and $e_{t_\imath}$ into global regression error controls for 
$\mathbb{E}^h[|Y^h_{t_{i}}-\hat{Y}^h_{t_{i}}|]$
and
$\mathbb{E}^h[|{\rho}^h_{t_{i}}-\hat{\rho}^h_{t_{i}}|]$
is more involved.
 We define the sequences of non-negative polynomials $(\PolyP{0}{i})_{i\geq 1}$, $(\PolyP{1}{i})_{i\geq 1}$, $(\PolyQ{0}{i})_{i\geq 1}$, and $(\PolyQ{1}{i})_{i\geq 1}$ by $\PolyP{0}{1} =\PolyQ{1}{1}= 1$, $\PolyP{1}{1} = \LamPhi/(1 - \alpha)$, $\PolyQ{0}{1} = 0$ and, for $i\ge 2$, 
\begin{align*}
     \PolyP{0}{i} & = \glb1+\glb1+\frac{\LamPhi}{1-\alpha}\grb x\grb \glb1+\glb1+\frac{2\LamPhi}{1-\alpha}\grb x\grb^{i-2},\\
     \PolyP{1}{i} & = \frac{2\LamPhi}{1-\alpha}\glb1+\glb1+\frac{\LamPhi}{1-\alpha}\grb x\grb\glb1+\glb1+\frac{2\LamPhi}{1-\alpha}\grb x\grb^{i-2},\\
     \PolyQ{0}{i} & = x\glb1+\glb1+\frac{2\LamPhi}{1-\alpha}\grb x\grb^{i-2},\\
     \PolyQ{1}{i} & = \frac{2\LamPhi x}{1-\alpha}\glb1+\glb1+\frac{2\LamPhi}{1-\alpha}\grb x\grb^{i-2}.
\end{align*}

\bt\label{t:induction}  With $\epsilon_{t_i} $ and $\thee_{t_i}$ as per \eqref{e:epsmu}, we have for every $\iidct \in\{0, \dots, \then\}$:
\beql{e:regerr_general_bound}
\left\{
\begin{aligned}
\mathbb{E}^h[|Y^h_{t_{\iidct}}-\hat{Y}^h_{t_{\iidct}}|] &\leq \sum_{i=1}^{\then-\iidct} \PolyP[\lambdadt]{0}{i}\sqrt{\mathbb{E}^h[\epsilon_{t_{\iidct+i-1}}^2]}+\PolyQ[\lambdadt]{0}{i}\sqrt{\mathbb{E}^h[\thee_{t_{\iidct+i-1}} ^2]}\\
\mathbb{E}^h[|{\rho}^h_{t_{\iidct}}-\hat{\rho}^h_{t_{\iidct}}|] &\leq \sum_{i=1}^{\then-\iidct} \PolyP[\lambdadt]{1}{i}\sqrt{\mathbb{E}^h[\epsilon_{t_{\iidct+i-1}}^2]}+\PolyQ[\lambdadt]{1}{i}\sqrt{\mathbb{E}^h[\thee_{t_{\iidct+i-1}}^2]}.
\end{aligned}
\right.
\eeql
\et
\proof 
By $\Lambda_f $ Lipschitz continuity of $f_{{k}} (t, x, y, \varrho)$ with respect to $y$ and $\varrho$, we have:
\[\mathbb{E}^h[|Y_{t_i}^h-\tilde{Y}_{t_i}^h|] \leq (1+\Lambda_f  \Delta t)\mathbb{E}^h[|Y^h_{t_{i+1}}-\hat{Y}^h_{t_{i+1}}|]+\Lambda_f \Delta t\mathbb{E}^h[|{\rho}^h_{t_{i+1}}-\hat{\rho}^h_{t_{i+1}}|]\]
and the triangular inequality yields
\beql{e:err_bound_Y-A}
\mathbb{E}^h[|Y_{t_i}^h-\hat{Y}_{t_i}^h|] \leq (1+\Lambda_f  \Delta t)\mathbb{E}^h[|Y^h_{t_{i+1}}-\hat{Y}^h_{t_{i+1}}|]+\mathbb{E}^h[|\hat{Y}_{t_i}^h-\tilde{Y}_{t_i}^h|]+\Lambda_f \Delta t\mathbb{E}^h[|{\rho}^h_{t_{i+1}}-\hat{\rho}^h_{t_{i+1}}|].
\eeql
By the $(1-\alpha)^{-1}$ Lipschitz continuity \eqref{e:esLip} of the expected shortfall, the $\LamPhi$ 
 Lipschitz continuity of $\Phi^h$, and an application of the triangular inequality, we have
{\allowdisplaybreaks
\begin{align*}
\mathbb{E}^h[|\rho^h_{t_{i}}-\tilde{\rho}^h_{t_{i}}|]\le & \E^h\glc\biggl|\mathbb{ES}^h_{t_{i}}[\Phi^h_{\tb_i 
}\big( Y ^{\pi} _{t_\cdot} +\sum_{\imath  < \cdot} f(t_\imath , \cX^h_{t_\imath }, Y^{\pi }_{t_{\imath+1} }, \rho^{\pi }_{t_{\imath+1} } )\Delta t ]\gra\\
&\gla- \mathbb{ES}^h_{t_{i}}[\Phi^h_{\tb_i 
}\big( \hatY ^{\pi} _{t_\cdot} +\sum_{\imath  < \cdot} f(t_\imath , \cX^h_{t_\imath },\hat{Y}^{\pi }_{t_{\imath+1} }, \hat{\rho}^{\pi }_{t_{\imath+1} } )\Delta t ]\biggr|\grc\\
{}_\text{($\ES^h_t$ is Lipschitz \eqref{e:esLip})}\le &\Afac \underbrace{\E^h\Biggl[\E^h_{t_i}}_{\E^h}\glc \biggl|\Phi^h_{\tb_i 
}\big( Y ^{\pi} _{t_\cdot} +\sum_{\imath  < \cdot} f(t_\imath , \cX^h_{t_\imath }, Y^{\pi }_{t_{\imath+1} }, \rho^{\pi }_{t_{\imath+1} } )\Delta t  \gra\\
& \gla\gla- \Phi^h_{\tb_i 
}\big( \hatY ^{\pi} _{t_\cdot} +\sum_{\imath  < \cdot} f(t_\imath , \cX^h_{t_\imath },\hat{Y}^{\pi }_{t_{\imath+1} }, \hat{\rho}^{\pi }_{t_{\imath+1} } )\Delta t\biggl|\grc\grc\\
{}_\text{($\Phi^h$ is Lipschitz \eqref{e:phihLip})}\le & \Afac[\LamPhi] \E^h \Biggl[\biggl|Y ^{\pi}_{ t_{(i+\them)\wedge\then}}+ Y ^{\pi}_{ t_i   }  -\sum_{\imath=i}^{(i+\them -1)\wedge(\then-1)} f(t_\imath , \cX^h_{t_\imath },{Y}^{\pi }_{t_{\imath+1} }, {\rho}^{\pi }_{t_{\imath+1} })\Delta t\\
& - \hatY ^{\pi}_{ t_{(i+\them)\wedge\then}}-\hatY ^{\pi}_{ t_i   }  +\sum_{\imath=i}^{(i+\them -1)\wedge(\then-1)} f(t_\imath , \cX^h_{t_\imath },\hat{Y}^{\pi }_{t_{\imath+1} }, \hat{\rho}^{\pi }_{t_{\imath+1} })\Delta t\biggr|\Biggr]\\
\le & \Afac[\LamPhi] \E^h\Biggl[|Y ^{\pi}_{ t_{(i+\them)\wedge\then}}- \hatY ^{\pi}_{ t_{(i+\them)\wedge\then}}| + |Y ^{\pi}_{ t_i   } - \hatY ^{\pi}_{ t_i   }|\\
& + \sum_{\imath=i}^{(i+\them -1)\wedge(\then-1)}|f(t_\imath , \cX^h_{t_\imath },Y^{\pi }_{t_{\imath+1} }, \rho^{\pi }_{t_{\imath+1} })-f(t_\imath , \cX^h_{t_\imath },\hat{Y}^{\pi }_{t_{\imath+1} }, \hat{\rho}^{\pi }_{t_{\imath+1} })|\Delta t\Biggr]\\
{}_\text{($f$ is Lipschitz \eqref{e:fLip})}\le & \Afac[\LamPhi]\Biggl[ \E^h[|Y ^{\pi}_{ t_{(i+\them)\wedge\then}}- \hatY ^{\pi}_{ t_{(i+\them)\wedge\then}}|\ + \E^h[|Y ^{\pi}_{ t_i   } - \hatY ^{\pi}_{ t_i   }|]\\
& + \lambdadt\sum_{\imath=i}^{(i+\them -1)\wedge(\then-1)}\glb\E^h[|Y^{\pi }_{t_{\imath+1} }-\hat{Y}^{\pi }_{t_{\imath+1} }| + \E^h[|\rho^{\pi }_{t_{\imath+1} }-\hat{\rho}^{\pi }_{t_{\imath+1} }|]\grb\Biggr].
\end{align*}}
Hence
\beql{e:err_bound_rho}
\begin{aligned}
\mathbb{E}^h[|{\rho}^h_{t_{i+1}}-\hat{\rho}^h_{t_{i+1}}|]  \leq & \mathbb{E}^h[|\tilde{\rho}^h_{t_{i+1}}-\hat{\rho}^h_{t_{i+1}}|] + \mathbb{E}^h[|\rho^h_{t_{i+1}}-\tilde{\rho}^h_{t_{i+1}}|]\\\leq & \mathbb{E}^h[|\tilde{\rho}^h_{t_{i+1}}-\hat{\rho}^h_{t_{i+1}}|] \\
&+\frac{\LamPhi}{1-\alpha}\left(\mathbb{E}^h[|{Y}^h_{t_{(i+\them+1)\wedge n}}-\hat{Y}^h_{t_{(i+\them+1)\wedge n}}|]+\mathbb{E}^h[|{Y}^h_{t_{i+1}}-\hat{Y}^h_{t_{i+1}}|]\right)\\
&+ \frac{{\LamPhi}\Lambda_f \Delta\,t}{1-\alpha}\sum_{\imath=i+2}^{(i+\them+1)\wedge(n-1)} \left(\mathbb{E}^h[|{Y}^h_{t_{\imath}}-\hat{Y}^h_{t_{\imath}}|]+\mathbb{E}^h[|{\rho}^h_{t_{\imath}}-\hat{\rho}^h_{t_{\imath}}|]\right).
\end{aligned}
\eeql
The inequalities \eqref{e:regerr_general_bound} are then shown by strong induction on $k$, which is detailed in Section \ref{s:induction}.~\finproof\\

 \noindent
The global controls \eqref{e:regerr_general_bound}, which by inspection of the proof are not far from sharp, reflect the geometrical accumulation of the local errors inherent to all dynamic programming based simulation/regression algorithms
(cf.~the comment following \citep[Eqn.~(VIII.3.13)]{gobet2016monte}). As explained in \citet[Section 3]{abbasturki:hal-01714747}, completely avoiding regressions would lead to multiply nested Monte Carlo, with one more level of nesting per pricing time step in particular. Highly multiply nested Monte Carlo is of course impractical, especially on realistic banking portfolios. Regression schemes are more practical, but the take-away message of Theorem \ref{t:induction} is that, as usual with such schemes simulation/regression schemes, the number of regression steps should be limited. This is why in particular we distinguish a finer simulation time grid from a coarser pricing (regression) time grid in our implementation\footnote{see https://github.com/BouazzaSE/NeuralXVA.}. From a practical error control viewpoint, the real usefulness of the twin Monte Carlo procedure is at the level of the computation of the $\sqrt{\mathbb{E}^h[ \epsilon^{2}_{t_i} ]}$, the way detailed in the end of Section \ref{ss:valid}.

\section{XVA Framework \label{sec:the_xva_case}}
\subsection{Continuous-Time Setup}
We consider a bank dealing financial derivatives
with multiple counterparties indexed by $c$, with default times $\tau^{(c)}$, where all portfolios are uncollateralized  
with zero recovery in the case of defaults (all assumed instantaneously settled)\footnote{for notational simplicity we assume no contractual cash flows between the bank and client $c$ at the exact time $\tau^{(c)}$.}.  
We denote by $\tmop{MtM}^{(c)}$ the aggregated mark-to-market process (counterparty-risk-free valuation) of
the portfolio of the bank with counterparty $c$. The bank, with risky funding spread process $\gamma^{(b)}$,
maintains capital at risk at the level of  an economic capital (EC) defined below as an
expected shortfall of the bank trading loss over one year, at a confidence level
$\alpha \in (\frac{1}{2},1)$. The bank is assumed perfectly hedged in terms of market risk\footnote{in the sense of \citet[Eqn.~(2.10)]{Crepey21}}, hence its trading loss reduces to the one of its CVA and FVA desks. 
For the sake of brevity in notation, we omit the discountings at the risk-free rate in the equations (in other terms, we use the risk-free asset as a num\'eraire), while preserving them in the numerical codes.
We assume a KVA risk premium at a hurdle rate $\therh >0$, i.e.~bank shareholders earn dividends at rate $\therh $ on their capital at risk. 
Finally we assume as in \citet*{CrepeyElie16} that the bank can use its capital 
as a risk-free funding source. 

As detailed in 
\citeN[Section A.2]{CrepeyHoskinsonSaadeddine2019} and \citeN[Theorem 6.1]{Crepey21}, this yields the following CVA (credit valuation
adjustment), FVA (funding valuation adjustment), EC (economic capital) and KVA
(capital valuation adjustment) 
equations, where $J^{(c)}_t=\indi{t< \tau^{(c)}}$ (so  $-dJ^{(c)}_t= \delta_{\tau^{(c)}} (dt) $, the Dirac measure at time $\tau^{(c)}$): For $t\le T$ (the final maturity of the derivative portfolio of the bank),
\beql{e:xvas}
 & \tmop{CVA}_t =  \mathbb{E}_t \left[\sum_c \int_t^T
  (\tmop{MtM}^{(c)}_s)^+ \delta_{\tau^{(c)}} (ds) \right] \\
&  \tmop{FVA}_t= \mathbb{E}_t \left[ \int_t^T \gamma^{(b)}_s  \left( \sum_c
  J^{(c)}_s \tmop{MtM}^{(c)}_s- \tmop{CVA}_s -
  \tmop{FVA}_s - \max (\tmop{EC}_s, \tmop{KVA}_s) \right)^+ ds
  \right]\\
 &  \tmop{KVA}_t = \mathbb{E}_t \left[ \int_t^T \therh  e^{- \therh  (s - t)} \max
  (\tmop{EC}_s, \tmop{KVA}_s) ds \right],
\eeql
where,  with $\tp=(t+1)\wedge T$,
\beql{e:EC}
& \tmop{EC}_t = \mathbb{E}\mathbb{S}_t
 [\loss_{\tp } - \loss_t] 
,
\eeql
in which the loss process $\loss$ satisfies, starting from $\loss_0=0$:
\beql{e:L}
  &d\loss_t =  \sum_c (\tmop{MtM}^{(c)}_t)^+ \delta_{\tau^{(c)}}  (dt)+d\tmop{CVA}_t + \\
&\qqq  \gamma^{(b)}_s  \left( \sum_c  J^{(c)}_t \tmop{MtM}^{(c)}_t
 - \tmop{CVA}_t - \tmop{FVA}_t - \max
  (\tmop{EC}_t, \tmop{KVA}_t) \right)^+ dt
  +
  d\tmop{FVA}_t .
\eeql 

All the random variables $J^{(c)}_t$ and $\tmop{MtM}^{(c)}_t$, as well the pre-default intensity $\gamma^{(b)}_t$ of the bank, are assumed to be $\sigma(\cX_t)$ measurable. Hence so is
$\tmop{CVA}_t$ as per the first line in \eqref{e:xvas}. 
The FVA and the KVA equations in \eqref{e:xvas} can then  be written in the form \eqref{BSDE},
for $l=2$ and
\beql{e:mapping}
 &Y_t= \left(\begin{aligned}&\tmop{FVA}_t,\\ &e^{-\therh  t}\tmop{KVA}_t\end{aligned}\right)\sp \phi=(0,0)^{\top},
\\
&f_{k}(t,x,(y_1,y_2)^{\top}, \varrho ) = \\& \left(\begin{aligned}& \gamma^{(b)}_k(t, x)\big(\sum_{c}J_k^{(c)}(t, x){\rm MtM}^{(c)}_k(t,x)-{\rm CVA}_k(t,x)-y_{1}-\max(\varrho,e^{\therh  t}y_{2})\big)^+\\ &\therh  \max(e^{-\therh  t} \varrho, y_{2})\end{aligned}\right) ,\\&
\qqq \mbox{
  where we denote ${Z}_k(t,x)=\mathbb{E}^h[Z_t|X_t=x, J_t=k]$,  for any process $Z$}, \\
& \Phi_{\overline{\cdot} }(M) =\int_{\cdot}^{\bar{\cdot} }  \left(  \sum_c (\tmop{MtM}^{(c)}_t)^+ \delta_{\tau^{(c)}} (dt)+ d\tmop{CVA}_t + dM^1_t \right)  \mbox{,  for any  }     M=(M^1, M^2)^{\top}\in\cS^2_2 
\eeql
(so, in this XVA setup, $\Phi$ only depends on $M$ via $M^1$).
{Note that, for $(Y,M)$ solving the ABSDE \eqref{BSDE} corresponding to the specification \eqref{e:mapping}, we have
\bel & 
 dM^{1}_t=d\tmop{FVA}_t+ \gamma^{(b)}_t  \left( \sum_c
  J^{(c)}_t  \tmop{MtM}^{(c)}_t- \tmop{CVA}_t -
   \tmop{FVA}_t - \max (\tmop{EC}_t, \tmop{KVA}_t) \right)^+ dt,
   \eel
   hence in view also of \eqref{e:L}:
 \beql{e:mappingbis}
 &  \theR_{\tp }(M) =\loss_{\tp }-\loss_t ,
   \eeql
which completes the connection between \eqref{e:mapping} and \eqref{e:xvas}--\eqref{e:L}.

\brem \label{rem:optimiXVA} In this XVA case, due to the special form \eqref{e:mappingbis} of $\theR_{\tp }(M)$, one does not need to maintain 
a full {\em process} $\hat{M}^{h}_{t_i}$ 
numerically as it appears in
\eqref{approxBSDEd} / Algorithm \ref{alg:explicit_backward_algo}.
It suffices to update iteratively in (decreasing) time $t_i$  the random variables $\loss_{\tb_i}-\loss_{t_i}$ or (in the Picard case) $\loss^{j}_{\tb_i}-\loss^{j}_{t_i}$ the way detailed in
\eqref{e:ECVA} and  \eqref{e:ECVAbis}.
\erem 
Note that the corresponding functional $\Phi$ satisfies in view of its formulation in \eqref{e:mapping}\footnote{cf.~\eqref{e:whirhointr}.}:
 \beql{e:Phim} &|\Phi(t;\mathbf{x},\mathbf{m})-\Phi(t;\mathbf{x},\mathbf{m}')|\le |\mathbf{m}_{\tp }- \mathbf{m}'_{\tp }| 
,\eeql 
so that Assumption \ref{hypgeneratorbis}(iii) holds with $\Lambda_{\Phi} =1$.  
Assumption \ref{hypgeneratorbis}(i) is readily checked. Assumption \ref{hypgeneratorbis}(ii) holds provided the process $\big( \gamma^{(b)}(t,\cX_t)(\sum_{c}J(t,\cX_t){\rm MtM}^{(c)}(t,\cX_t)\big)$ is in $\cH^2$, e.g.~for $ \gamma^{(b)}$ bounded and $\big( {\rm MtM}^{(c)}(t,\cX_t)\big)$
in $\cH^2$ for each client $c$.
  }
 
In view of the first line in \eqref{e:xvas},
the CVA process can be estimated by nonparametric least squares regression in space, at each grid pricing time $t_i$, based on Monte Carlo paths
of the forward process $\cX$\footnote{or its time-discretized version $\cX^h$, cf.~Section \ref{sec:xva_explicit_scheme}.}. 
The exercise is made delicate, however, by the hybrid nature of $\cX$, which includes both diffusive (market risk) and discrete (default risk) components. This difficulty is solved by the hierarchical simulation scheme
 of \citeN{AbbasturkiCrepeySaadeddine20},
 whereby many default trajectories are simulated conditional on each simulated trajectory of the market. Based on this, the CVA process is treated hereafter as a given (already estimated) process.
 
The FVA, EC, and KVA equations
(with CVA now assumed exogenously given in \eqref{e:xvas}--\eqref{e:L})
are challenging due to their
coupling via the loss process $\loss$. 
However, this coupling can be overcome by a
combination of time discretization schemes and (or not) Picard
iterations, the way presented in the more general context
of Section \ref{sec:approx} for the explicit scheme and 
hereafter
in the XVA setup for both the explicit and the implicit/Picard scheme.   
 
\brem\label{rem:redvar}
In practice, for variance reduction purposes, we use for CVA computations a default intensities based reformulation of the CVA, instead of its definition based on default indicators in \eqref{e:xvas}. Still we do use the hierarchical simulation scheme of \citeN{AbbasturkiCrepeySaadeddine20}, simulating many default paths given each realization of the diffusion processes, in order to help with learnings where we do not have the convenience of using default intensities, i.e. given the presence of default terms in the loss \eqref{e:L}, which occur nonlinearly in EC computations, and of default terms under the $(\cdot)^+$ in the FVA computations.
\erem

{\def\thec{\gamma}\def\thec{c}
\def\Deltat{h}\def\Deltat{\Delta t}
To alleviate the notation, we assume a uniform time step $ \Deltat_{i+1}=\Deltat $, with  $1/\Deltat$ chosen as a positive integer $m$ and $\tp_\thei =t_{(\thei +m) \wedge n}$, and
we skip all the $\hat{\cdot}$ and $\cdot^h$ later in this section (with a last exception to mention that
$\LamPhi=1$, as can be readily checked from below the same way as we checked $\Lambda_{\Phi}=1$ in \eqref{e:Phim}).
By least squares (resp.~quantile) regressions below, we always mean \textit{neural network} least squares (resp.~quantile) regressions conducted  against all risk factors (market risk factors and client default indicators) at each decreasing time
  $t_\thei $, the way detailed in Section \ref{ss:algos}. All regressions and quantile regressions are implemented using a neural network of one hidden layer with 38 neurons\footnote{for experimentation with the network architecture see \citeN[Section 5.1, Figure 2]{CrepeyHoskinsonSaadeddine2019}.}.

\subsection{Explicit Simulation/Regression XVA Scheme\label{sec:xva_explicit_scheme}}

Here we use the following discretization:
$\tmop{CVA}_{t_n}=\tmop{FVA}_{t_n}=\tmop{KVA}_{t_n}=0$ and, for $\thei =n-1\cdots 0,$
\begin{align}
& \tmop{CVA}_{t_\thei } =   \mathbb{E}_{t_\thei } \Biggl[ \sum_c  \sum_{\thei  \leq \imath  \leq n
  - 1} (\tmop{MtM}^{(c)}_{t_{\imath  +1 }})^+  \ind_{\{ t_\imath  < \tau^{(c)}
  \leq t_{\imath  + 1} \}} \Biggr], \nonumber\\ 
&  \tmop{FVA}_{t_\thei } =   \mathbb{E}_{t_\thei } \Biggl[ \tmop{FVA}_{t_{\thei  + 1}} + \nonumber\\&\left.\quad
\Deltat  \gamma^{(b)}_{t_{i  }}  \left( \sum_c \tmop{MtM}^{(c)}_{t_{i  }}
  \ind_{\{ \tau^{(c)} > t_{i  } \}} - \tmop{CVA}_{t_{i  }} -
  \tmop{FVA}_{t_{\thei  + 1}} - \max (\tmop{EC}_{t_{\thei  + 1}}, \tmop{KVA}_{t_{i +
  1}}) \right)^+ \right], \nonumber\\
&  \tmop{KVA}_{t_\thei } =   \exp (- \therh  \Deltat ) \mathbb{E}_{t_\thei }
  [\tmop{KVA}_{t_{\thei  + 1}} + \therh  \Deltat  \max (\tmop{EC}_{t_{\thei  + 1}},
  \tmop{KVA}_{t_{\thei  + 1}})],\nonumber\\\nonumber
 & \tmop{EC}_{t_\thei }  
  =  \mathbb{E}\mathbb{S}_{t_\thei }
  [L_{\tp _{i}} -
 L_{t_\thei }], \mbox{ where }
\\&\quad\quad
\label{e:ECVA}
L_{t_{\thei  + 1}} - L_{t_\thei }=  \tmop{CVA}_{t_{\thei  + 1}} - \tmop{CVA}_{t_\thei } +
  (\tmop{MtM}^{(c)}_{t_{i }})^+  \ind_{\{ t_\thei  < \tau^{(c)} \leq
  t_{\thei  + 1} \}}+ \tmop{FVA}_{t_{\thei  + 1}} - \tmop{FVA}_{t_\thei }+ \\&\qqq \Deltat  \gamma^{(b)}_{t_{i  }}  \left( \sum_c \tmop{MtM}^{(c)}_{t_{i  }} \ind_{\{ \tau^{(c)}
  > t_{i  } \}} - \tmop{CVA}_{t_{i }} - \tmop{FVA}_{t_{\thei  + 1}} - \max
  (\tmop{EC}_{t_{\thei  + 1}}, \tmop{KVA}_{t_{\thei  + 1}}) \right)^+ .\nonumber
\end{align}
We recognize the explicit scheme of Algorithm \ref{alg:explicit_backward_algo} applied to $Y=(\FVA,\KVA)$. This scheme naturally lifts the coupling visible in \eqref{e:xvas}--\eqref{e:L} between $\tmop{EC}$, $\tmop{KVA}$, and
$\tmop{FVA}$. Specifically, assuming that all the
$\tmop{XVA}_{t_{\thei+1} } $ and $\tmop{EC}_{t_{\thei+1} } $ have already been learned, we compute:
\begin{enumerate}
  \item $\tmop{CVA}_{t_i}$, by a least-squares regression of $\sum_c \sum_{\imath 
  \leq i \leq n - 1} (\tmop{MtM}^{(c)}_{t_{\imath  }})^+  \ind_{\{ t_\imath  <
  \tau^{(c)} \leq t_{\imath  + 1} \}}$\footnote{or equivalent variance-reduced cash flows formulated in terms of the default \textit{intensities} as explained in Remark \ref{rem:redvar}.} against all risk factors at time $t_\thei $;

    \item $\tmop{FVA}_{t_\thei } $, through a least-squares regression of
  \bel&\tmop{FVA}_{t_{\thei  + 1}} +\\&  \quad \Deltat  \gamma^{(b)}_{t_{i  }} \left( \sum_c
  \tmop{MtM}^{(c)}_{t_{i  }} \ind_{\{ \tau^{(c)} > t_{i } \}} -
  \tmop{CVA}_{t_{i  }} - \tmop{FVA}_{t_{\thei  + 1}} - \max (\tmop{EC}_{t_{i +
  1}}, \tmop{KVA}_{t_{\thei  + 1}}) \right)^+\eel against all risk factors at time
  $t_\thei $;
  
  \item $\tmop{KVA}_{t_\thei }$, through a least-squares regression of $$\exp (- \therh 
  \Deltat )  (\tmop{KVA}_{t_{\thei  + 1}} + \therh  \Deltat  \max (\tmop{EC}_{t_{\thei  + 1}},
  \tmop{KVA}_{t_{\thei  + 1}}))$$ against all risk factors at time $t_\thei $;
    \item $\tmop{EC}_{t_\thei }$, through quantile regression of $L_{\tb_i} - L_{t_\thei }$ followed by a least-squares regression to deduce the expected
  shortfall as detailed in Section \ref{ss:algos}, both regressions involving all the
  risk factors at time $t_\thei $.
\end{enumerate} 

\paragraph{A Posteriori Twin Monte Carlo Validation Procedure} 
Let
\beql{e:eeps}\epsilon^{fva}_{t_i}=\tmop{FVA}_{t_i}-\tilde{\tmop{FVA}}_{t_i} \sp\epsilon^{kva}_{t_i}=
\tmop{KVA}_{t_i}-
\tilde{\tmop{KVA}}_{t_i}, \eeql
where $(\tilde{\tmop{FVA}}_{t_i},e^{-r t_i}\tilde{\tmop{KVA}}_{t_i})$ corresponds to
\eqref{e:Yrt} in the $( \tmop{FVA}_{t_i},e^{-r t_i}\tmop{KVA}_{t_i})$ case \eqref{e:mapping}.
The twin error estimation \eqref{e:twins} applied to $(\tmop{FVA}_{t_i},e^{-r t_i} \tmop{KVA}_{t_i})$ (component-wise as detailed in the proof of Lemma \ref{lem:twin}) yields  
\begin{align}
\mathbb{E}^h &[(\epsilon^{fva}_{t_i})^2] =  \mathbb{E}^h\left[(\tmop{FVA}_{t_i})^2\right.\nonumber\\ &\label{e:twin_fva}-\tmop{FVA}_{t_i}\left(\tmop{FVA}^{(1)}_{t_{i+1}}+f^1(t_i, \cX^{\pi}_{t_i}, \hat{Y}^{\pi,(1)}_{t_{i+1}},\hat{\rho}_{t_i+1}^{h,(1)})\Delta {t}\right.\left.+ \tmop{FVA}^{(2)}_{t_{i+1}}+f^1(t_i, \cX^{\pi}_{t_i}, \hat{Y}^{\pi,(2)}_{t_{i+1}},\hat{\rho}_{t_i+1}^{h,(2)})\Delta {t}\right)\\ &\left.+\glb\tmop{FVA}^{(1)}_{t_{i+1}}+f^1(t_i, \cX^{\pi}_{t_i}, \hat{Y}^{\pi,(1)}_{t_{i+1}},\hat{\rho}_{t_i+1}^{\pi,(1)})\Delta {t})^{\top}(\tmop{FVA}^{(2)}_{t_{i+1}}+f^1(t_i, \cX^{\pi}_{t_i}, \hat{Y}^{\pi,(2)}_{t_{i+1}},\hat{\rho}_{t_i+1}^{\pi,(2)})\Delta {t}\grb\right],\nonumber    
\end{align}
where $\tmop{FVA}^{(1)}_{t_{i+1}}$ and $\tmop{FVA}^{(2)}_{t_{i+1}}$ are independent conditioned on $\cX^{\pi}_{t_i}$, and $f^1$ is the first component of the vector function $f$. Substituting $\exp^{-rt_i}\tmop{KVA}_{t_i}$ instead of $\tmop{FVA}_{t_i}$ and $f^2$ instead of  $f^1$ in \eqref{e:twin_fva}, one obtains $e^{-2r t_i}\times$the twin error estimation $\mathbb{E}^h[(\epsilon^{kva}_{t_i})^2]$ for $ \tmop{KVA}_{t_i}$.

\subsection{Implicit/Picard Simulation/Regression XVA Scheme}\label{sec:xva_implicit_scheme}

We define and compute the CVA as in the explicit scheme. For the rest of the XVAs, we
introduce Picard iterations echoing \eqref{contraction_system-picard}, starting from $\tmop{EC}^{0}\equiv0$ followed by, for increasing $j\ge 1$:
 $\tmop{FVA}^{\thej }_{t_n}=\tmop{KVA}^{\thej }_{t_n}=0$ and, for $\thei =n-1\cdots 0,$
\begin{align}
 & \tmop{FVA}^{\thej }_{t_\thei } =  \mathbb{E}_{t_\thei } \Biggl[ \tmop{FVA}^{\thej }_{t_{i
  + 1}} + \nonumber\\  & \quad\left.
\Deltat  \gamma^{(b)}_{t_\thei }  \left( \sum_c \tmop{MtM}^{(c)}_{t_\thei }
  \ind_{\{ \tau^{(c)} > t_\thei  \}} - \tmop{CVA}_{t_\thei } - \tmop{\widetilde{FVA}}^{(\thej  -
  1)}_{t_\thei } - \max (\tmop{EC}^{\thej  - 1}_{t_\thei }, \tmop{\widetilde{KVA}}_{t_\thei }^{\thej  - 1})
  \right)^+ \right]\nonumber\\
 &  \tmop{KVA}^{\thej }_{t_\thei } = \exp (- \therh  \Deltat ) \mathbb{E}_{t_\thei }
  [\tmop{KVA}^{\thej }_{t_{\thei  + 1}} + \therh  \Deltat  \max (\tmop{EC}^{\thej-1 }_{t_{i }}, \tmop{\widetilde{KVA}}^{ \thej-1  }_{t_{i  }})]\nonumber\\
 & \tmop{EC}^{\thej }_{t_\thei } =  \mathbb{E}\mathbb{S}_{t_\thei }
  [L^{\thej }_{\tp _i} - L^{\thej }_{t_\thei }], \text{where}\label{e:ECVAbis}
\\
  \nonumber
 &\quad\quad L^{\thej }_{t_{\thei  + 1}} - L^{\thej }_{t_\thei } = \tmop{CVA}_{t_{\thei  + 1}} -
  \tmop{CVA}_{t_\thei } + (\tmop{MtM}^{(c)}_{t_{i }})^+  \ind_{\{ t_\thei  <
  \tau^{(c)} \leq t_{\thei  + 1} \}}+ \tmop{FVA}^{\thej }_{t_{\thei  + 1}} - \tmop{FVA}^{\thej }_{t_\thei } +\\
  & \qqq\Deltat 
  \gamma^{(b)}_{t_\thei }\left( \sum_c \tmop{MtM}^{(c)}_{t_\thei } \ind_{\{
  \tau^{(c)} > t_\thei  \}} - \tmop{CVA}_{t_\thei } - \tmop{\widetilde{FVA}}^{\thej  - 1}_{t_\thei } - \max
  (\tmop{EC}^{\thej  - 1}_{t_\thei }, \tmop{\widetilde{KVA}}^{\thej  - 1}_{t_\thei }) \right)^+\nonumber ,
\end{align}
where we set
\beql{e:Yinitter}
&\tmop{\widetilde{FVA}}^{\thej -1}_{t_{i}}=\ind_{j=1} \tmop{FVA}^{\thej   }_{t_{i+1}} +\ind_{j\ge 2}\tmop{FVA}^{\thej -1  }_{t_{i}} \sp \tmop{\widetilde{KVA}}^{\thej -1}_{t_{i}}=\ind_{j=1} \tmop{KVA}^{\thej   }_{t_{i+1}} +\ind_{j\ge 2}\tmop{KVA}^{\thej -1  }_{t_{i}}.\eeql

In this scheme the coupling between $\tmop{EC}$, $\tmop{KVA}$, and
$\tmop{FVA}$ is removed by the Picard iterations in $j$.  Specifically, assuming that all the
$\tmop{XVA}^{j-1}_{t_{\thei } } $ (for $j\ge 2$), $\tmop{XVA}^{j}_{t_{\thei+1} } $, and $\tmop{EC}^{j-1}_{t_{\thei } } $ have already been learned, we compute: 
\begin{enumerate}
  \item $\tmop{FVA}^{\thej }_{t_\thei } $, by least-squares regression of
 \bel&\tmop{FVA}^{\thej }_{t_{\thei  + 1}} + \\&\quad \Deltat  \gamma^{(b)}_{t_\thei }  \left( \sum_c
  \tmop{MtM}^{(c)}_{t_\thei } \ind_{\{ \tau^{(c)} > t_\thei  \}} -
  \tmop{CVA}_{t_\thei } - \tmop{\widetilde{FVA}}^{ \thej -1 }_{t_\thei } - \max (\tmop{EC}^{ \thej -1 }_{t_\thei },
  \tmop{KVA}_{t_\thei }^{ \thej -1 }) \right)^+\eel against the risk factors at time $t_\thei $;
  
    \item $\tmop{KVA}^{\thej }_{t_\thei }$, through a least-squares regression of
  $$\exp (- \therh\Deltat )  (\tmop{{KVA}}^{ \thej  }_{t_{\thei  + 1}} + \therh  \Deltat  \max
  (\tmop{EC}^{ \thej -1 }_{t_{i  }}, \tmop{\widetilde{KVA}}^{ \thej -1 }_{t_{i  }}))$$ against all
  risk factors at time $t_\thei $;
  
  \item $\tmop{EC}_{t_\thei }^{\thej }$, through quantile regression of $L^{\thej }_{\tb_i} - L^{\thej }_{t_\thei }$ followed by a least-squares
  regression to deduce the expected
  shortfall as detailed in Section \ref{ss:algos}, both regressions involving all the
  risk factors at time $t_\thei $.
\end{enumerate}

}

\section{XVA Numerical Benchmark}\label{ss:numr}
 \afterpage{
 \begin{table}[htbp]
\centering
\begin{tabular}{lrrrrr}
\toprule
{} &  $j=1$ &  $j=2$ &   $j=3$ &   $j=4$ &    Explicit \\
\midrule
$h=\frac{T}{2^5}$ &  463.279938 &  433.832031 &  434.391296 &  433.753998 &  434.65167 \\
$h=\frac{T}{2^6}$ &  461.329926 &  433.141876 &  434.036011 &  433.835052 &  433.60974 \\
$h=\frac{T}{2^7}$  &  461.031097 &  432.506531 &  433.631531 &  431.789215 &  433.18683 \\
$h=\frac{T}{2^8}$  &  460.326050 &  433.123596 &  431.992859 &  432.098328 &  434.29538 \\
\bottomrule
\end{tabular}
\caption{$\FVA_0$ under the Picard iteration scheme of Section \ref{sec:xva_implicit_scheme}
vs.~the explicit scheme.}
\label{tab:standard_picard_vs_explicit_t0}
\end{table}
}

 \afterpage{
\begin{figure}[htbp]
  \includegraphics[width=\linewidth]{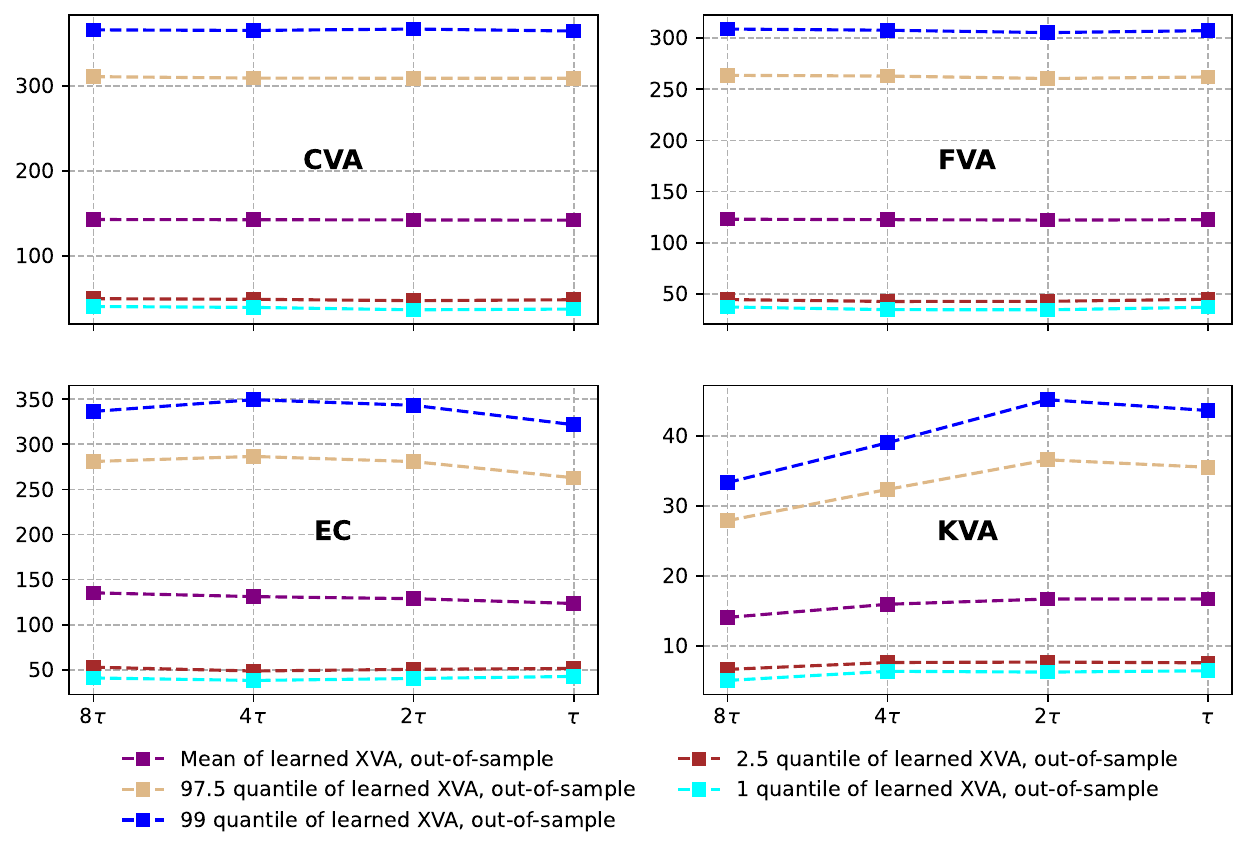}
  \caption{Mean and quantiles of CVA, FVA, KVA and EC learned by the explicit scheme at $t=\frac{T}{2}$ for different sizes of the time step.}
  \label{fig:convergence_all_slice}
\end{figure}
 }
\afterpage{\begin{figure}[htbp]
  \includegraphics[width=\linewidth]{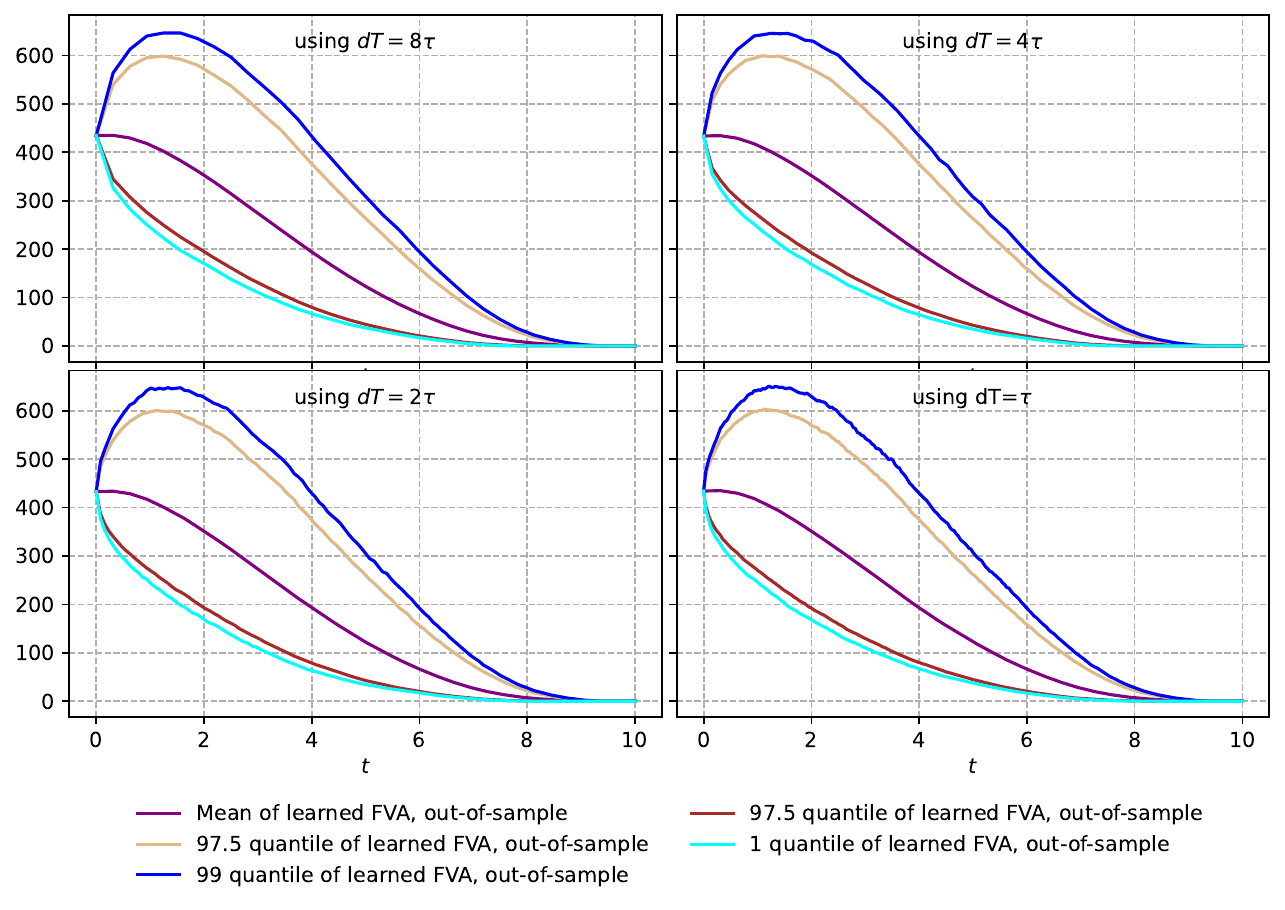}
  \caption{FVA profiles using an explicit scheme.}
  \label{fig:explicit_fva_profiles}
\end{figure}}

\afterpage{\begin{figure}[htbp]
  \includegraphics[width=\linewidth]{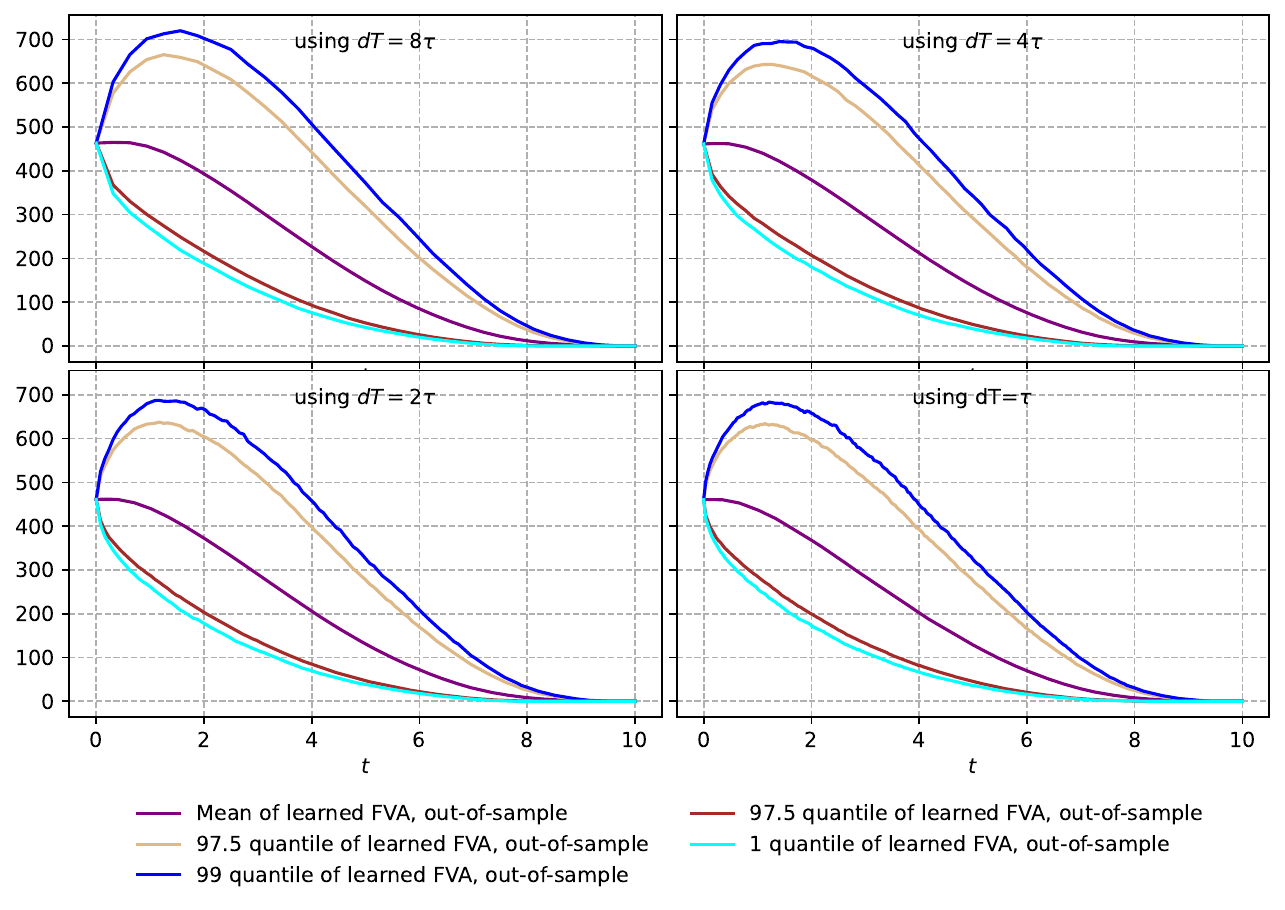}
  \caption{FVA profiles obtained after $j=1$ Picard iteration for the implicit scheme.}
  \label{fig:implicit_fva_profiles_01}
  \includegraphics[width=\linewidth]{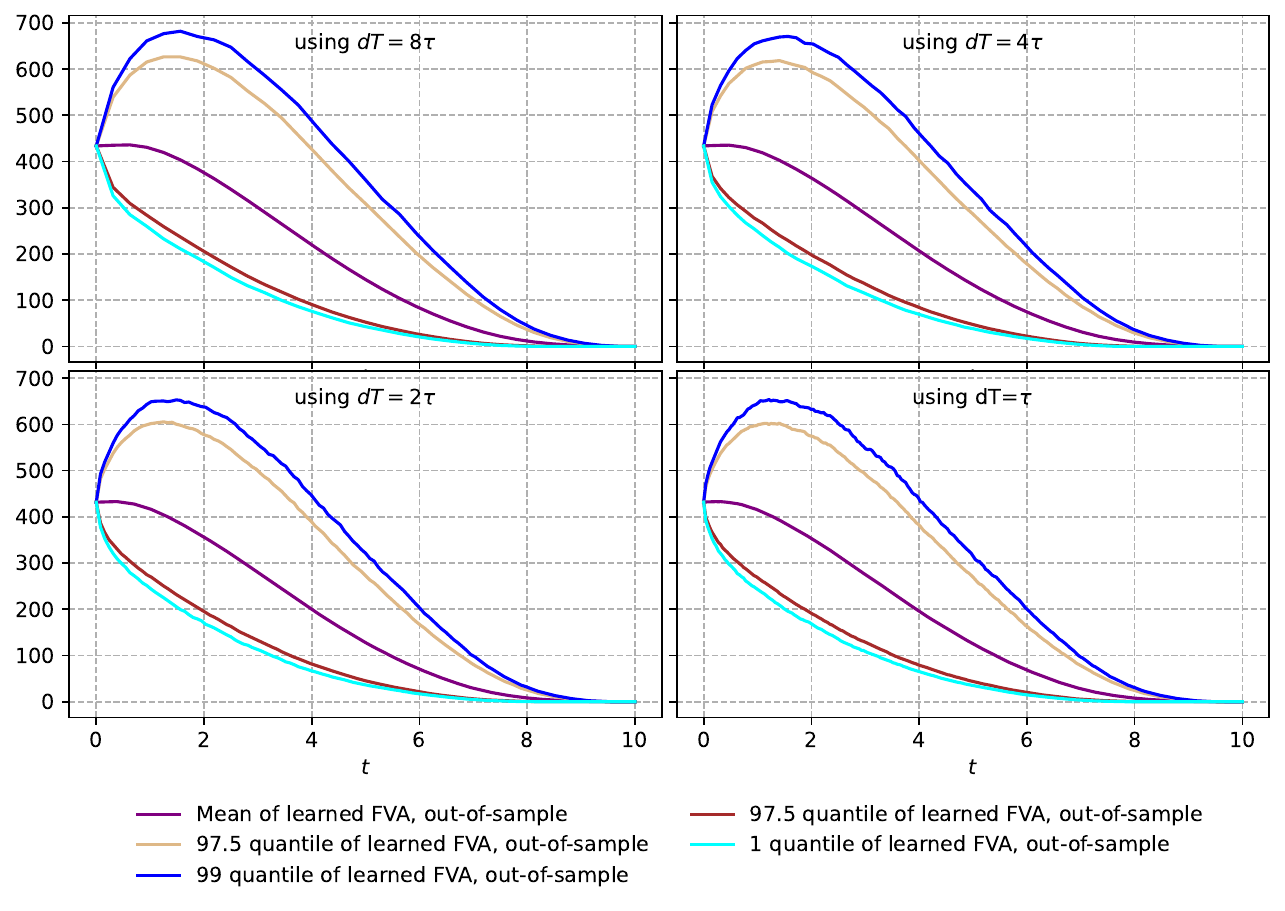}
  \caption{FVA profiles obtained after $j=4$ Picard iteration for the implicit scheme.}
  \label{fig:implicit_fva_profiles_04}
\end{figure}}

\afterpage{\begin{figure}[htbp]
  \includegraphics[width=\linewidth]{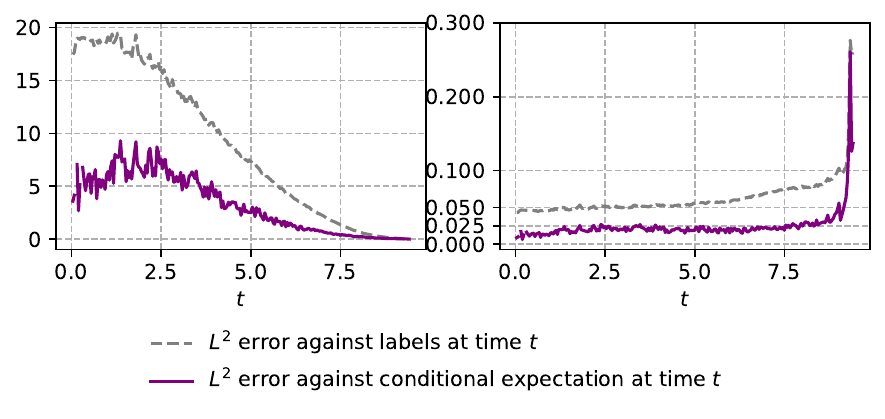}
  \caption{Local regression errors $\sqrt{\mathbb{E}^h[(\epsilon^{fva}_{t_i})^2]}$ (solid purple) vs.~$L^2$ training losses (dashed grey) \blu{for $\theta=10$}. Left panel: raw errors. Right panel: errors normalized at each time step $t_i$ by the $L^2$ norm of $\hat\FVA^h_{t_i}$.}
  \label{fig:local_l2_err_via_twin_fva}
\end{figure}}

\afterpage{\begin{figure}[htbp]
  \includegraphics[width=\linewidth]{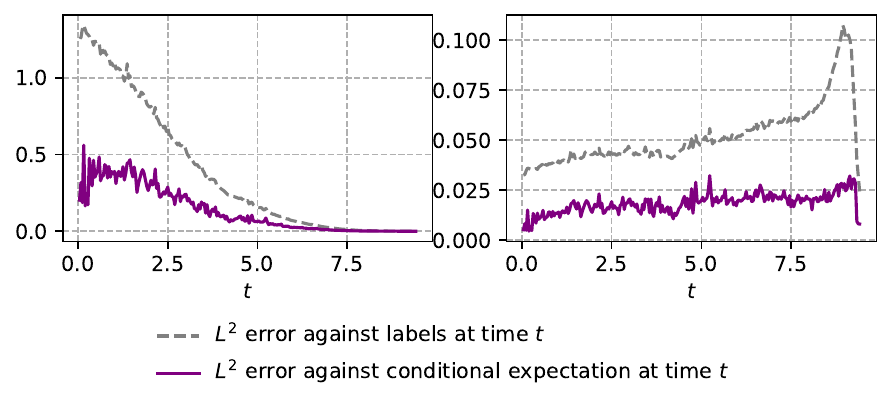}
  \caption{Local regression errors $\sqrt{\mathbb{E}^h[(\epsilon^{kva}_{t_i})^2]}$ (solid purple)  vs.~$L^2$ training losses (dashed grey). Left panel: raw errors.  Right panel: errors normalized at each time step $t_i$ by the $L^2$ norm of $\hat\KVA^h_{t_i}$.}
  \label{fig:local_l2_err_via_twin_kva}
\end{figure}}

\afterpage{\begin{figure}[htbp]
  \includegraphics[width=\linewidth]{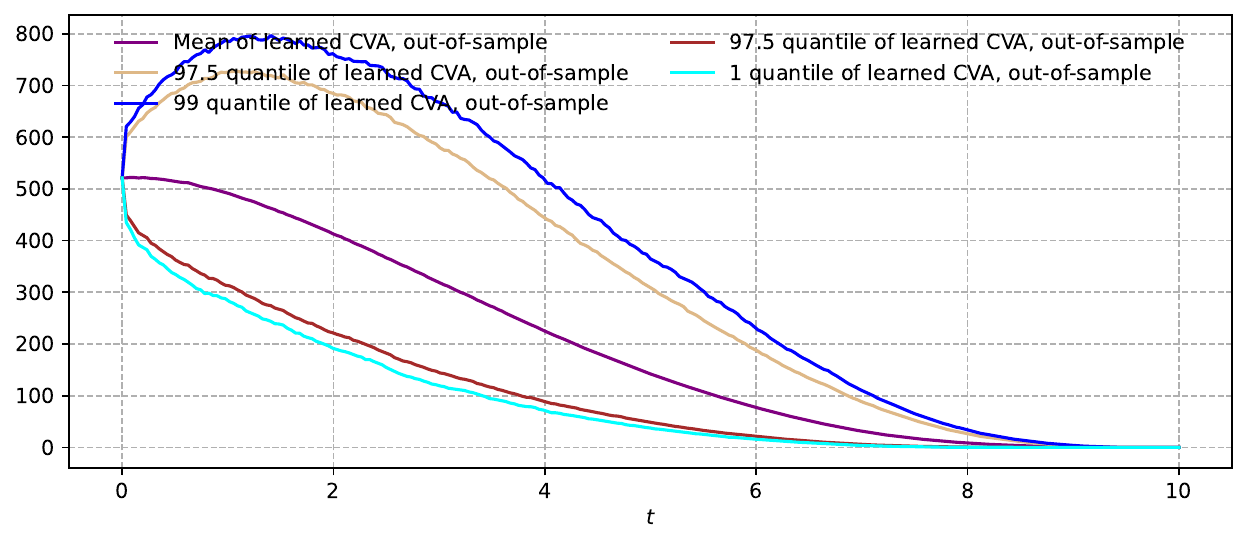}
  \caption{CVA profiles using an explicit scheme and a fine time discretization ($\theta=10$)}
  \label{fig:cva_profiles_fine}
  \includegraphics[width=\linewidth]{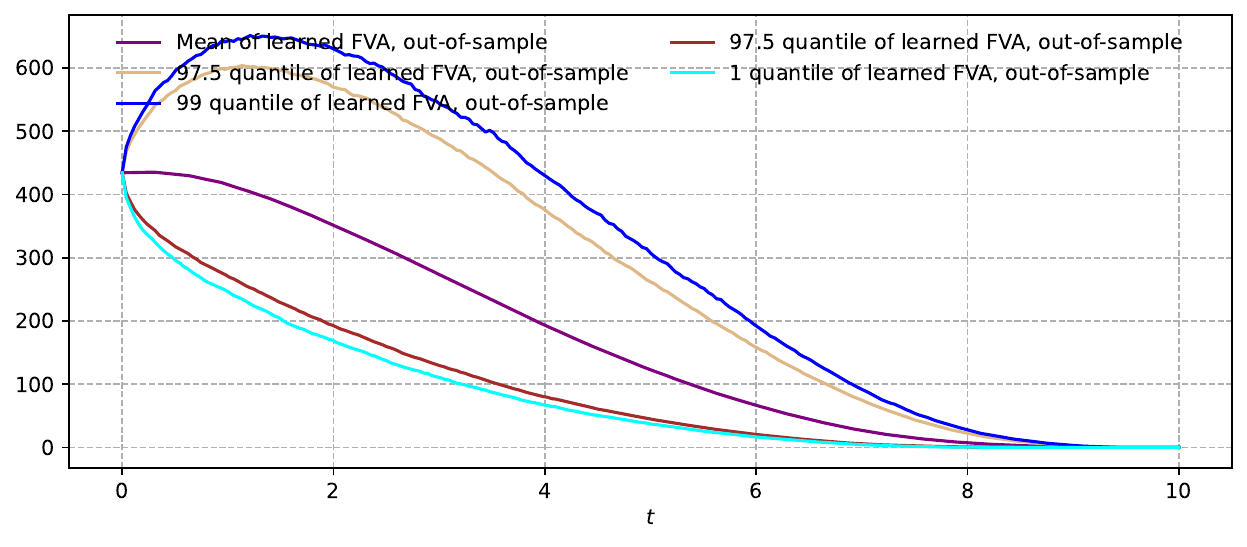}
  \caption{FVA profiles using an explicit scheme and a fine time discretization ($\theta=10$)}
  \label{fig:fva_profiles_fine}
  \includegraphics[width=\linewidth]{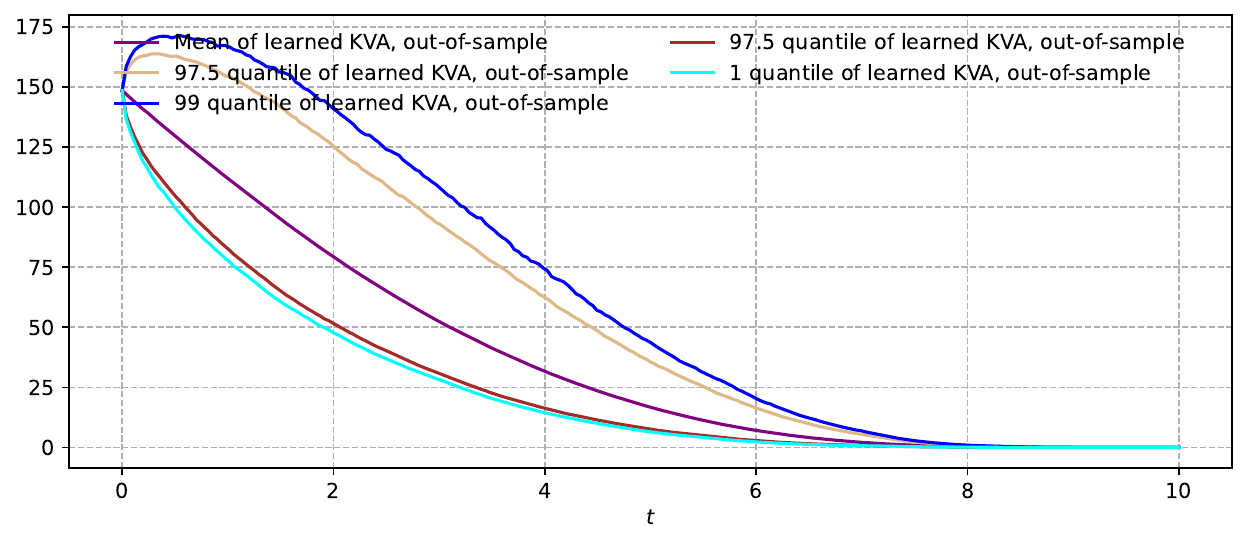}
  \caption{KVA profiles using an explicit scheme and a fine time discretization ($\theta=10$)}
  \label{fig:kva_profiles_fine}
\end{figure}}
 
 \afterpage{\begin{figure}[htbp]
   \includegraphics[width=\linewidth]{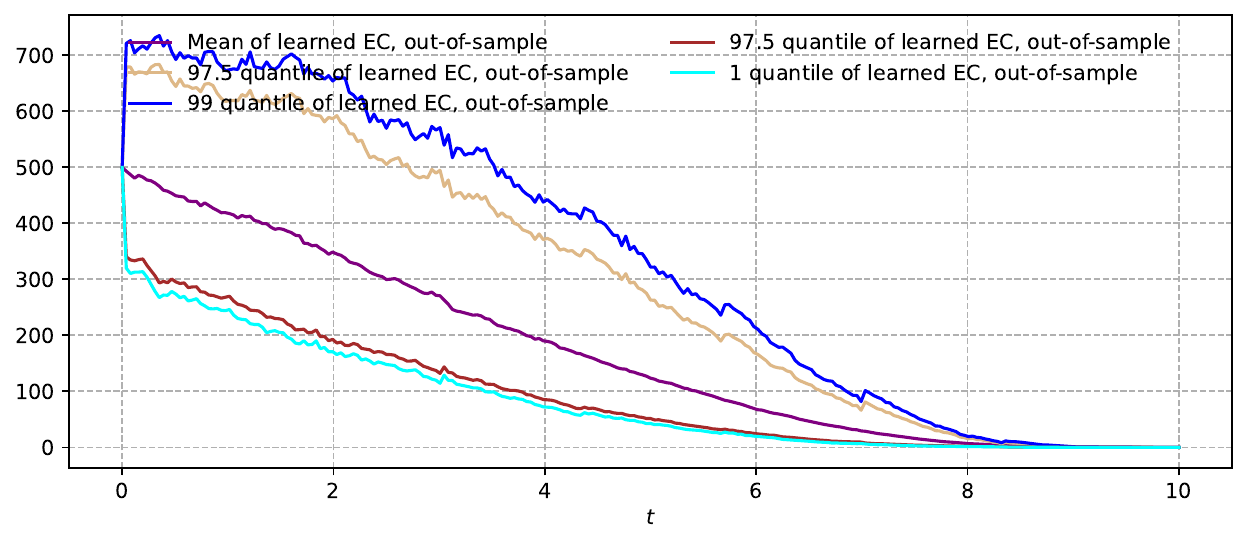}
   \caption{EC profiles using an explicit scheme and a fine time discretization ($\theta=10$)}
   \label{fig:ec_profiles_fine}
 \end{figure}}

For our numerical experiments, we assume $10$ economies, each represented by a short-rate process with Vasicek dynamics, $9$ cross-currency rate processes with log-normal dynamics, 
$8$ counterparties each with a stochastic default intensity process following CIR dynamics, adding up to $28$ diffusive risk factors (when accounting for the spread of the bank which is also assumed to be driven by a CIR process) and 8 default indicator processes.
We consider a portfolio of $100$ interest rate swaps, all at-par at time $0$. The characteristics of the swaps (notional, maturity, counterparty and currency) are drawn randomly. In particular, the maturities of the swaps range between $0.9375$ and $T=10$ years. For all modeling and implementation details, as well as numerical values of all the parameters, the reader is referred to https://github.com/BouazzaSE/NeuralXVA.

\subsection{Benchmarking the Explicit versus the Picard Scheme\label{ss:bench}}

For the purpose of empirical time discretization analysis\footnote{see end of Section \ref{ss:td}.}, we consider multiple time discretizations $(\pi^{(\theta)})_\theta$ such that
\[
\pi^{(\theta)} = \{t^{(\theta)}_i := i\frac{T}{2^\theta}, i=0, \dots,2^\theta\}.
\]
In Table \ref{tab:standard_picard_vs_explicit_t0} and Figures  
\ref{fig:convergence_all_slice}--\ref{fig:implicit_fva_profiles_04},
we experimented for $\theta\in\{5,6,7,8\}$ ($\tau$ in the figures then corresponds to $\frac{T}{2^8}$).  We can thus see the convergence of the Picard iterations for the FVA process, using 
the implicit $(\FVA,\KVA)$ scheme of Section 
\ref{sec:xva_implicit_scheme},
 toward a solution visually very close, already for $j=4$, to the one 
 provided by the explicit scheme 
of Algorithm \ref{alg:explicit_backward_algo}, aka Section \ref{sec:xva_explicit_scheme}. However, because of the multiple Picard iterations, the Picard scheme comes at the cost of $j=3$-$4$ times longer computations  (for reference, the explicit scheme takes 7mins 48secs using the finest time grid,  i.e.~$\theta=8$, on an NVidia A100 GPU).
These results mutually validate the explicit and the implicit/Picard scheme against each other, while showing that the Picard scheme is not competitive with respect to the explicit scheme in terms of computation times.

\subsection{A Posteriori Error Validation of the Explicit Scheme}

{The solid purple curves in Figures \ref{fig:local_l2_err_via_twin_fva}  and \ref{fig:local_l2_err_via_twin_kva} exhibit the local regression errors $\sqrt{\mathbb{E}^h[(\epsilon^{fva}_{t_i})^2]}$ and $\sqrt{\mathbb{E}^h[(\epsilon^{kva}_{t_i})^2]}$ of the $(\FVA_{t_i}, e^{-r t_i}\KVA_{t_i})$ scheme\footnote{cf.~\eqref{e:eeps}.}  \blu{for a fine time discretization $\theta=10$.}
The dashed grey curves represent the corresponding $L^2$ training losses. The comparison between the grey and purple curves shows the benefit of the a posteriori Monte Carlo local regression error estimates \eqref{e:twins} with respect to the $L^2$ training losses that would be used as a naive and overconservative  error estimate.} These and the above numerical results confort us in the choice of the explicit scheme 
as the reference pathwise XVA numerical scheme.
Our final plots in
Figures \ref{fig:cva_profiles_fine}--\ref{fig:ec_profiles_fine} show profiles for respectively the CVA, FVA, KVA and EC when using the explicit scheme \blu{with $\theta=10$}.  

\section{Conclusion\label{sec:concl}} 

 The main contributions of the paper are the ABSDE explicit scheme of Algorithm  \ref{alg:explicit_backward_algo}, which dominates the implicit/Picard scheme in terms of computation times (for mutually validating XVA numbers) in the XVA case study of Section \ref{ss:numr},  and the
a posteriori error control of Section \ref{s:valid} for the explicit scheme.

The recent and fastly growing machine learning literature on the solution of high-dimensional nonlinear BSDEs/PDEs contains, at least, two branches. The first one, in the line of \citeN{EHanJentzen17}, consists in learning together the time-0 value of the solution and the gradient-process of the latter through a single learning task. Examples in the XVA space include \citeN{HenryLabordere17rm} or \citeN{gnoatto2020deep}. 
The former reference provides insightful PDE views
on the CVA and MVA. The second reference computed the XVAs and their model state sensitivities simultaneously. 
However, the stylized equations of \citeN{HenryLabordere17rm} are 
quite far
from actual XVA equations. The XVA equations of \citeN{gnoatto2020deep} are less unrealistic, but they are still restricted to computations at the level of one client of the bank and they are accordingly handled by reduction of filtration in the line of \citeN{Crepey13a1}, so that the default times ultimately disappear from the equations.
Such an approach does not leverage to several clients, whose default times enter the XVA equations of the bank in a nonlinear fashion (and therefore have to be simulated as we do).
The second stream of literature, that includes \citeN{HurePhamWarin19} or the present paper, learns the solution time step after time step (in backward time), much like classical dynamic programming algorithms, except that 
machine learning techniques are used for solving the local equations which then arise at each successive decreasing pricing time step. 

In the case of our ABSDE XVA equations, the first global approach would not be viable on realistic problems stated at the portfolio level due to the significant memory requirement for the corresponding global training task. 
In contrast, the second local approach benefits from a synergy between the successive local training tasks involved: as all our XVA equations are endowed with zero terminal conditions, 
 the variance of the cash flows (``regressands") entering the successive learning tasks as input data at the decreasing pricing time steps increases progressively (time step after time step), whereas the variance of the features (``regressors"), i.e.~of the risk factors, decreases.   
As a result, the difficulty of the training tasks gradually increases throughout the course of the algorithm\footnote{In order to see this in a simplified setup, consider linear regression instead of neural networks. When close to $t=0$, the variances of the regressors tend to $0$, which leads to ill-conditioned covariance matrices, while the variance of the regressands increases, which makes the regression even more unstable.}. But the next training task also benefits from all previous ones, via the use of the weights trained at a time step as initialization for the weights at the next time step. This is probably one of the reasons behind the 
robustness of our approach\footnote{provided the hierarchical simulation technique of \citeN{AbbasturkiCrepeySaadeddine20} and the best practice  
risk measure estimates of \citeN{BarreraCrepeyGobetSaadeddine} are used, as done in our numerics.},
while a global approach would fail on unsolvable memory occupation issues.

From an algorithmic viewpoint, work in progress
aims at demonstrating how the XVA simulation/regression framework of this paper can be leveraged to also encompass XVA sensitivities. Note that AAD sensitivities computational techniques
\`a la
\citet*{baydin2018automatic,savine2018modern} 
are not a viable alternative in a pathwise XVA setup where the pathwise XVA metrics are already the output of optimization training procedures: AAD sensitivity techniques can only be available in more rudimentory XVA frameworks. 
From a mathematical viewpoint, the establishment of a Feynman-Kac representation for the solution of the limiting ABSDE \eqref{BSDE}, as well as the study of the time-consistency of our explicit scheme,  
are interesting open issues.

 \appendix

\section{End of the Proof of Theorem \ref{t:induction}}\label{s:induction}

\bl\label{lem:re_poly_identities}
For all $i\geq 1$:
{\rm\hfill\break \textbf{(i)}}  $(1+x)\PolyP{0}{i}+x \PolyP{1}{i} \le \PolyP{0}{i+1}$.
 {\rm\hfill\break \textbf{(ii)}} $(1+x)\PolyQ{0}{i}+x \PolyQ{1}{i} \le \PolyQ{0}{i+1}$.\\
If $i \le \them$, then:
 {\rm\hfill\break \textbf{(iii)}} $\PolyP{0}{i}+x\sum_{\isumi=1}^{i-1}\glb\PolyP{0}{\isumi}+\PolyP{1}{\isumi}\grb \leq (1-\alpha) \PolyP{1}{i}/\LamPhi$.
 {\rm\hfill\break \textbf{(iv)}} $\PolyQ{0}{i}+x\sum_{\isumi=1}^{i-1}\glb\PolyQ{0}{\isumi}+\PolyQ{1}{\isumi}\grb \leq (1-\alpha) \PolyQ{1}{i}/\LamPhi$.\\
If $i \ge \them + 1$, then:
{\rm\hfill\break \textbf{(v)}} $\PolyP{0}{i}+ \PolyP{0}{i-\them}+x\sum_{\isumi=i-\them}^{i-1}\glb\PolyP{0}{\isumi}+\PolyP{1}{\isumi}\grb \leq (1-\alpha) \PolyP{1}{i}/\LamPhi$.
{\rm\hfill\break \textbf{(vi)}} $\PolyQ{0}{i}+ \PolyQ{0}{i-\them}+x\sum_{\isumi=i-\them}^{i-1}\glb\PolyQ{0}{\isumi}+\PolyQ{1}{\isumi}\grb \leq (1-\alpha) \PolyQ{1}{i}/\LamPhi$.
\el

\proof
It is straightforward to verify that the polynomials $\PolyP{0}{i}, \PolyP{1}{i}, \PolyQ{0}{i}, \PolyQ{1}{i}$ are generated by $\PolyP{0}{i}=1, \PolyP{1}{i}=\LamPhi/(1-\alpha), \PolyQ{0}{i}=0, \PolyQ{1}{i}=1$, and the following recursive relations, for $i\ge 2$:
\begin{align}
    & \PolyP{0}{i} = (1 + x)\PolyP{0}{i - 1} + x\PolyP{1}{i - 1},\label{eq:poly_gen_p2}\\
        & \PolyP{1}{i} = \Afac[2\LamPhi\PolyP{0}{i}], \label{eq:poly_gen_p1}\\
     & \PolyQ{0}{i} = (1 + x)\PolyQ{0}{i - 1} + x\PolyQ{1}{i - 1},\label{eq:poly_gen_q2}
\\         & \PolyQ{1}{i} = \Afac[2\LamPhi\PolyQ{0}{i}] \label{eq:poly_gen_q1}.
\end{align}
With \eqref{eq:poly_gen_p2} and \eqref{eq:poly_gen_q2}, \textbf{(i)} and \textbf{(ii)} are proven and sharp.  Equation \eqref{eq:poly_gen_p2} for $i\ge 2$ can be rewritten as 
\begin{equation*}
    \PolyP{0}{i + 1} - \PolyP{0}{i} = x\glb\PolyP{1}{i} + \PolyP{0}{i}\grb\sp i\ge 1,
\end{equation*}
which leads to the handy relation
\begin{equation}
    x\sum_{\isumi=1}^{i-1} \glb\PolyP{0}{\isumi} + \PolyP{1}{\isumi}\grb= \PolyP{0}{i} - \PolyP{0}{1}\sp i\ge 1.
    \label{eq:poly_sum_eq}
\end{equation}
Inserting \eqref{eq:poly_sum_eq} into \eqref{eq:poly_gen_p1}, $\PolyP{1}{i}$ can be rewritten, for $i\ge 2$, as
\begin{align}
    (1 - \alpha)\PolyP{1}{i}/\LamPhi & = \PolyP{0}{i} + \PolyP{0}{i} + \PolyP{0}{1} - \PolyP{0}{1}\\
    & = \PolyP{0}{i} + x\sum_{\isumi=1}^{i-1} \glb\PolyP{0}{\isumi} + \PolyP{1}{\isumi}\grb + \PolyP{0}{1}\label{eq:poly_sum_eq2}\\
    & \ge \PolyP{0}{i} + x\sum_{\isumi=1}^{i-1} \glb\PolyP{0}{\isumi} + \PolyP{1}{\isumi}\grb.\label{eq:poly_sum_ineq}
\end{align}
Inequality \eqref{eq:poly_sum_ineq} is also true for $i= 1$, hence \textbf{(iii)} is proven. For $i\ge m + 1$, inserting \eqref{eq:poly_sum_eq} into \eqref{eq:poly_sum_eq2} leads to
\begin{align*}
    (1 - \alpha)\PolyP{1}{i}/\LamPhi = & \PolyP{0}{i} + x\sum_{\isumi=1}^{i-1} \glb\PolyP{0}{\isumi} + \PolyP{1}{\isumi}\grb + \PolyP{0}{1}\\ &+ \underbrace{\PolyP{0}{i-\them} - \PolyP{0}{1} - x\sum_{\isumi=1}^{i - m - 1}\glb\PolyP{0}{\isumi} + \PolyP{1}{\isumi}\grb}_0\nonumber\\
    = & \PolyP{0}{i} + \PolyP{0}{i-\them} + x\sum_{\isumi=i-\them}^{i-1}\glb\PolyP{0}{\isumi}+\PolyP{1}{\isumi}\grb,
\end{align*}
which proves \textbf{(v)} and indicates that it is sharp. It is worth noting that the proof requires only the recursive relations \eqref{eq:poly_gen_p1} and \eqref{eq:poly_gen_p2}, but not $\PolyP{0}{1}$ and $\PolyP{1}{1}$. As the polynomials $Q$ satisfy the same recursive relations as the polynomials $P$, \textbf{(iv)} and \textbf{(vi)} are also proven.~\finproof

Let $\Bound{0}{\iidct}$ and $\Bound{1}{\iidct}$ be defined, for $\iidct\le\then -1$, as
\begin{align}
    \Bound{0}{\iidct} & = \sum_{\isumo=1}^{\then-\iidct}\glb\PolyP[\lambdadt]{0}{\isumo}\eep{\iidct +\isumo - 1} + \PolyQ[\lambdadt]{0}{\isumo}\ee{\isumo + \iidct - 1}\grb;\label{eq:induct_assump_0}\\
    \Bound{1}{\iidct} & = \sum_{\isumo=1}^{\then-\iidct}\glb\PolyP[\lambdadt]{1}{\isumo}\eep{\iidct +\isumo - 1} + \PolyQ[\lambdadt]{1}{\isumo}\ee{\isumo + \iidct - 1}\grb.\label{eq:induct_assump_1}
\end{align}
With Lemma~\ref{lem:re_poly_identities}, we prove the theorem \ref{t:induction} by strong backward induction. Using \eqref{e:err_bound_Y-A} and \eqref{e:err_bound_rho}, we find $\EEP{\then-1} \le \eep{\then-1} = \Bound{0}{\then-1}$, and hence $\EE{\then-1} \le \Afac[\LamPhi] \eep{\then-1} + \ee{\then-1}=\Bound{1}{\then-1}$.
We assume that for each $i\ge k$ the errors $\EEP{\isumo}$ and $\EE{\isumo}$ are bounded by $\Bound{0}{\iidct}$ and $\Bound{1}{\iidct}$ respectively. 
Then, we want to prove, using this induction assumption, that $\EEP{\iidct - 1}$ and $\EE{\iidct - 1}$ are bounded by $\Bound{0}{\iidct - 1}$ and $\Bound{1}{\iidct - 1}$ respectively. The error of $\EEP{\iidct - 1}$ is bounded using \eqref{e:err_bound_Y-A}. By inserting \eqref{eq:induct_assump_0} into  \eqref{e:err_bound_Y-A}, we obtain
\begin{align*}
    \EEP{\iidct - 1} \le & (1 + \lambdadt)\Bound{0}{\iidct} + \lambdadt \Bound{1}{\iidct} + \eep{\iidct - 1}\\
    = & (1 + \lambdadt) \sum_{\isumo=1}^{\then - \iidct} \glb\PolyP[\lambdadt]{0}{\isumo}\eep{\iidct + \isumo - 1} + \PolyQ[\lambdadt]{0}{\isumo}\ee{\iidct + \isumo - 1}\grb\\
    & + \lambdadt \sum_{\isumo=1}^{\then-\iidct} \glb\PolyP[\lambdadt]{1}{\isumo}\eep{\iidct + \isumo - 1} + \PolyQ[\lambdadt]{1}{\isumo}\ee{\iidct + \isumo - 1}\grb\\
    & + \eep{\iidct - 1}\\
    = & \sum_{\isumo=2}^{\then-\iidct + 1} \underbrace{\glb(1 + \lambdadt)\PolyP[\lambdadt]{0}{\isumo-1} + \lambdadt\PolyP[\lambdadt]{1}{\isumo-1}\grb}_{\le \PolyP{0}{i} \text{ due to Lemma~\ref{lem:re_poly_identities}(i)}}\eep{\iidct + \isumo - 2}\\
    & + \sum_{\isumo=2}^{\then - \iidct + 1} \underbrace{\glb(1 + \lambdadt)\PolyQ[\lambdadt]{0}{\isumo-1} + \lambdadt\PolyQ[\lambdadt]{1}{\isumo-1}\grb}_{\le \PolyQ{0}{i}\text{ due to Lemma~\ref{lem:re_poly_identities}(ii)}}\ee{\iidct + \isumo - 2}\\
    & + \eep{\iidct - 1}\\
    \le & \sum_{\isumo=1}^{\then-\iidct + 1}\glb\PolyP[\lambdadt]{0}{\isumo}\eep{\iidct +\isumo - 2} + \PolyQ[\lambdadt]{0}{\isumo}\ee{\iidct + \isumo - 2}\grb\\
    = & \Bound{0}{\iidct - 1}.
\end{align*}
It is worth noting that the bound is obtained using not only Lemma~\ref{lem:re_poly_identities}, but also the value of $\PolyP{0}{1}$ and $\PolyQ{0}{1}$.

Applying the same idea to \eqref{e:err_bound_rho}, the bound of $\EE{i}$, we have
\begin{align}
    \EE{\iidct-1} \le & \ee{\iidct - 1} + \Afac[\LamPhi] \glb\Bound{0}{\iidct - 1} + \Bound{0}{(\iidct+ \them - 1) \land \then} + \lambdadt \sum_{\isumi = \iidct}^{(\iidct +\them - 1)\land (\then -1)}\glb\Bound{0}{\isumi} + \Bound{1}{\isumi}\grb\grb.
     \label{eq:EE_origin}
\end{align}
Depending on the relation between $ \iidct +\them$ and $\then$, we have two cases. 
\paragraph{Case $\iidct+ \them> \then  $}
In this case, \eqref{eq:EE_origin} reduces to
\begin{align*}
    \EE{\iidct-1} \le & \ee{\iidct - 1} + \Afac[\LamPhi] \glb\Bound{0}{\iidct - 1} + \lambdadt \sum_{\isumi = \iidct}^{\then - 1}\glb\Bound{0}{\isumi} + \Bound{1}{\isumi} \grb\grb\\
    = & \Afac[\LamPhi] \glb\sum_{\isumo=1}^{\then - \iidct + 1}\PolyP[\lambdadt]{0}{\isumo}\eep{\iidct + \isumo - 2}+\PolyQ[\lambdadt]{0}{\isumo}\ee{\iidct + \isumo - 2}\grb\\
    & + \Afac[\LamPhi\lambdadt] \sum_{\isumi = \iidct}^{\then - 1}\sum_{\isumo=1}^{\then-\isumi}\glb\left(\PolyP[\lambdadt]{0}{\isumo} + \PolyP[\lambdadt]{1}{\isumo}\right)\eep{\isumi + \isumo - 1} \gra\\
    & \gla+ \left(\PolyQ[\lambdadt]{0}{\isumo} + \PolyQ[\lambdadt]{1}{\isumo}\right)\ee{\isumi + \isumo - 1}\grb + \ee{\iidct - 1}.
\end{align*}
Notice that for any two real sequences $(a_i)_{i\geq 1}$ and $(b_i)_{i\geq 1}$ we have:
\begin{align}
\sum_{\isumi=\iidct}^{\then - 1}\sum_{\isumo=1}^{\then - \isumi} a_\isumo b_{\isumi + \isumo -1} &= \sum_{\isumi=1}^{\then - \iidct} \sum_{\isumo=1}^{\then - \iidct - \isumi +1} a_\isumo b_{\iidct+\isumi+\isumo-2} = \sum_{\isumi=1}^{\then - \iidct} \sum_{\isumo=\isumi+1}^{\then - \iidct +1} a_{\isumo-\isumi} b_{\iidct+\isumo-2}\nonumber\\
&= \sum_{\isumo=2}^{\then - \iidct +1} \glb\sum_{\isumi=1}^{\isumo-1} a_{\isumo-\isumi}\grb b_{\iidct+\isumo-2} = \sum_{\isumo=2}^{\then - \iidct +1} \glb\sum_{\isumi=1}^{\isumo-1} a_{\isumi}\grb b_{\iidct+\isumo-2}.
\label{eq:ab_sum}
\end{align}
Using this relation, the order of the summation can be changed and the double sum terms become
\begin{align*}
    \sum_{\isumo=2}^{\then - \iidct +1} \glb\sum_{\isumi=1}^{\isumo-1} \glb\PolyP[\lambdadt]{0}{\isumi}+\PolyP[\lambdadt]{1}{\isumi}\grb \eep{\iidct+\isumo-2}+\sum_{\isumi=1}^{\isumo-1} \glb\PolyQ[\lambdadt]{0}{\isumi}+\PolyQ[\lambdadt]{1}{\isumi}\grb \ee{\iidct+\isumo-2}\grb.
\end{align*}
The error of $\EE{i}$ is thus bounded by
\begin{align*}
    &\sum_{\isumo=2}^{\then -\iidct + 1} \underbrace{\Afac[\LamPhi] \glb\PolyP[\lambdadt]{0}{\isumo} + \lambdadt\sum_{\isumi=1}^{\isumo-1}\glb\PolyP[\lambdadt]{0}{\isumi} + \PolyP[\lambdadt]{1}{\isumi}\grb\grb}_{\le \PolyP[\lambdadt]{1}{\isumo} \text{ due to Lemma~\ref{lem:re_poly_identities}(iii) } } \eep{\iidct + \isumo - 2}\\
   & +  \sum_{\isumo=2}^{\then -\iidct + 1} \underbrace{\Afac[\LamPhi] \glb\PolyQ[\lambdadt]{0}{\isumo} + \lambdadt\sum_{\isumi=1}^{\isumo-1}\glb\PolyQ[\lambdadt]{0}{\isumi} + \PolyQ[\lambdadt]{1}{\isumi}\grb\grb}_{\le \PolyQ[\lambdadt]{1}{\isumo} \text{ due to Lemma~\ref{lem:re_poly_identities}(iv) } } \ee{\iidct + \isumo - 2}\\
    & + \Afac[\LamPhi] \underbrace{\PolyP[\lambdadt]{0}{1}}_{=1} \eep{\iidct - 1} + \left(\Afac[\LamPhi] \underbrace{\PolyQ[\lambdadt]{0}{1}}_{=0} + 1\right)\ee{\iidct - 1}\\
    \le & \sum_{\isumo=1}^{\then-\iidct + 1}\glb\PolyP[\lambdadt]{1}{\isumo}\eep{\iidct +\isumo - 2} + \PolyQ[\lambdadt]{1}{\isumo}\ee{\iidct + \isumo - 2}\grb\\
    = & \Bound{1}{\iidct - 1}.
\end{align*}

\paragraph{Case $\iidct+ \them \le \then $}
In this case, \eqref{eq:EE_origin} reduces to
\begin{align*}
    \EE{\iidct-1} \le &\ee{\iidct - 1} + \Afac[\LamPhi]\glb \Bound{0}{\iidct - 1} + \Bound{0}{\iidct+ \them - 1} + \lambdadt \sum_{\isumi = \iidct}^{\iidct +\them - 1}\glb\Bound{0}{\isumi} + \Bound{1}{\isumi}\grb\grb\\
    = & \ee{\iidct - 1} + \Afac[\LamPhi]\glb\sum_{\isumo=1}^{\then -\iidct + 1}\glb\PolyP[\lambdadt]{0}{\isumo}\eep{\iidct + \isumo - 2} + \PolyQ[\lambdadt]{0}{\isumo}\ee{\iidct + \isumo - 2}\grb\gra\\ &\gla+ \sum_{\isumo=1}^{\then - \iidct + 1 - \them}\glb\PolyP[\lambdadt]{0}{\isumo}\eep{\iidct + \isumo - 2 + \them} + \PolyQ[\lambdadt]{0}{\isumo}\ee{\iidct + \isumo - 2+\them}\grb\grb\\
    & + \Afac[\LamPhi\lambdadt] \sum_{\isumi = \iidct}^{\iidct +\them - 1}\sum_{\isumo=1}^{\then - \isumi}\glb\glb\PolyP[\lambdadt]{0}{\isumo} + \PolyP[\lambdadt]{1}{\isumo}\grb\eep{\isumi + \isumo - 1} \gra\\
    & \gla+ \left(\PolyQ[\lambdadt]{0}{\isumo} + \PolyQ[\lambdadt]{1}{\isumo}\grb\ee{\isumi + \isumo - 1}\grb\\
    = & \Afac[\LamPhi] \glb\sum_{\isumo=1}^{\then -\iidct + 1}\glb\PolyP[\lambdadt]{0}{\isumo}\eep{\iidct + \isumo - 2} + \PolyQ[\lambdadt]{0}{\isumo}\ee{\iidct + \isumo - 2}\grb\gra\\ &+ \underbrace{\sum_{\isumo=\them+1}^{\then - \iidct + 1}}_{=\sum_{\isumo=1}^{\then - \iidct + 1}\indi{\isumo \ge \them + 1}}\glb\PolyP[\lambdadt]{0}{\isumo-\them}\eep{\iidct + \isumo - 2} + \PolyQ[\lambdadt]{0}{\isumo-\them}\ee{\iidct + \isumo - 2}\grb\Biggr)\\
    & + \Afac[\LamPhi\lambdadt] \glb\sum_{\isumi = \iidct}^{\then - 1} - \sum_{\isumi = \iidct + \them}^{\then - 1}\grb\sum_{\isumo=1}^{\then - \isumi}\glb\glb\PolyP[\lambdadt]{0}{\isumo} + \PolyP[\lambdadt]{1}{\isumo}\grb\eep{\isumi + \isumo - 1} \gra\\
    & \gla+ \left(\PolyQ[\lambdadt]{0}{\isumo} + \PolyQ[\lambdadt]{1}{\isumo}\grb\ee{\isumi + \isumo - 1}\grb + \ee{\iidct - 1}.    
\end{align*}
For any two real sequences $(a_i)_{i\geq 1}$ and $(b_i)_{i\geq 1}$, using \eqref{eq:ab_sum}, we have:
\begin{align}
\glb\sum_{\isumi = \iidct}^{\then - 1} - \sum_{\isumi = \iidct + \them}^{\then - 1}\grb\sum_{\isumo=1}^{\then - \isumi} a_\isumo b_{\isumi + \isumo -1} &=  \sum_{\isumo=2}^{\then - \iidct +1} \glb\sum_{\isumi=1}^{\isumo-1} a_{\isumi}\grb b_{\iidct+\isumo-2} -  \sum_{\isumo=2}^{\then - \iidct +1-\them} \glb\sum_{\isumi=1}^{\isumo-1}a_{\isumi}\grb b_{\iidct+\isumo-2+\them}\nonumber\\
&= \sum_{\isumo=2}^{\then-\iidct+1} \left(\sum_{\isumi=1}^{\isumo-1} a_\isumi\right) b_{\iidct+\isumo-2} - \sum_{\isumo=\them+2}^{\then-\iidct+1} \left(\sum_{\isumi=1}^{\isumo-\them-1} a_\isumi\right) b_{\iidct + \isumo-2}\nonumber\\
& = \sum_{\isumo=2}^{\then-\iidct+1}\glb \sum_{\isumi=1}^{\isumo-1} - \indi{\isumo\ge\them + 1}\sum_{\isumi=1}^{\isumo - \them -1} \grb a_\isumi b_{\iidct + \isumo-2}.
\label{eq:ab_sum2}
\end{align}
Note that we can use $\indi{\isumo\ge\them + 1}$ instead of the expected $\indi{\isumo\ge\them + 2}$ in the last line,
because, when $\isumo = \them + 1$, $\sum_{\isumi=1}^{\isumo - \them -1} = 0$. Using \eqref{eq:ab_sum2}, the double sum terms in next-to-last line above can be written as
\begin{align*}
\begin{split}
    &\Afac[\LamPhi\lambdadt] \sum_{\isumo=2}^{\then-\iidct+1}\glb \sum_{\isumi=1}^{\isumo-1} - \indi{\isumo\ge\them + 1}\sum_{\isumi=1}^{\isumo - \them -1} \grb \glb\glb\PolyP[\lambdadt]{0}{\isumi} + \PolyP[\lambdadt]{1}{\isumi}\grb\eep{\iidct + \isumo - 2} \gra\\
    & \gla+ \glb\PolyQ[\lambdadt]{0}{\isumi} + \PolyQ[\lambdadt]{1}{\isumi}\grb\ee{\iidct + \isumo - 2}\grb,
\end{split}
\end{align*}
and the bound for $\EE{\iidct - 1}$ becomes

\newcommand{\glbb}{\left(\rule{0em}{8mm}\right.}
\newcommand{\grbb}{\left.\rule{0em}{8mm}\right)}
{\allowdisplaybreaks
\begin{align*}
    & \sum_{\isumo=2}^{\then-\iidct+1} \Afac[\LamPhi]\glbb\PolyP[\lambdadt]{0}{\isumo} + \indi{\isumo\ge \them + 1}\PolyP[\lambdadt]{0}{\isumo-\them}\\&\qqq +  \lambdadt\left(\sum_{\isumi=1}^{\isumo-1}-\indi{\isumo\ge \them+1}\sum_{\isumi=1}^{\isumo-\them-1}\right)\left(\PolyP[\lambdadt]{0}{\isumi}+\PolyP[\lambdadt]{1}{\isumi}\right)\grbb\eep{\iidct+\isumo-2}\\
    + & \sum_{\isumo=2}^{\then-\iidct+1}\Afac[\LamPhi]\glbb\PolyQ[\lambdadt]{0}{\isumo} + \indi{\isumo\ge \them + 1}\PolyQ[\lambdadt]{0}{\isumo-\them}\\&\qqq + \left(\sum_{\isumi=1}^{\isumo-1}-\indi{\isumo\ge \them+1}\sum_{\isumi=1}^{\isumo-\them-1}\right)\left(\PolyQ[\lambdadt]{0}{\isumi}+\PolyQ[\lambdadt]{1}{\isumi}\right)\grbb\ee{\iidct+\isumo-2}\\
    + & \Afac[\LamPhi] \PolyP[\lambdadt]{0}{1}\eep{\iidct - 1} + \left(\Afac[\LamPhi] \PolyQ[\lambdadt]{0}{1} + 1\right)\ee{\iidct - 1}\\
    =& \sum_{\isumo=\them + 1}^{\then-\iidct+1} \underbrace{\Afac[\LamPhi]\glb\PolyP[\lambdadt]{0}{\isumo} + \PolyP[\lambdadt]{0}{\isumo-\them} + \lambdadt\sum_{\isumi=\isumo -\them}^{\isumo-1}\left(\PolyP[\lambdadt]{0}{\isumi}+\PolyP[\lambdadt]{1}{\isumi}\right)\grb}_{\le \PolyP[\lambdadt]{1}{\isumo} \text{ due to Lemma~\ref{lem:re_poly_identities}(v)}}\eep{\iidct+\isumo-2}\\
    & + \sum_{\isumo=\them + 1}^{\then-\iidct+1} \underbrace{\Afac[\LamPhi]\glb\PolyQ[\lambdadt]{0}{\isumo} + \PolyQ[\lambdadt]{0}{\isumo-\them} + \lambdadt\sum_{\isumi=\isumo -\them}^{\isumo-1}\left(\PolyQ[\lambdadt]{0}{\isumi}+\PolyQ[\lambdadt]{1}{\isumi}\right)\grb}_{\le \PolyQ[\lambdadt]{1}{\isumo} \text{ due to Lemma~\ref{lem:re_poly_identities}(vi)}}\eep{\iidct+\isumo-2}\\
    & + \sum_{\isumo=2}^{\them} \underbrace{\Afac[\LamPhi]\glb\PolyP[\lambdadt]{0}{\isumo} + \lambdadt\sum_{\isumi=1}^{\isumo-1}\left(\PolyP[\lambdadt]{0}{\isumi}+\PolyP[\lambdadt]{1}{\isumi}\right)\grb}_{\le \PolyP[\lambdadt]{1}{\isumo} \text{ due to Lemma~\ref{lem:re_poly_identities}(iii)}}\eep{\iidct+\isumo-2}\\
    & + \sum_{\isumo=2}^{\them} \underbrace{\Afac[\LamPhi]\glb\PolyQ[\lambdadt]{0}{\isumo} + \lambdadt\sum_{\isumi=1}^{\isumo-1}\left(\PolyQ[\lambdadt]{0}{\isumi}+\PolyQ[\lambdadt]{1}{\isumi}\right)\grb}_{\le \PolyQ[\lambdadt]{1}{\isumo} \text{ due to Lemma~\ref{lem:re_poly_identities}(iv)}}\eep{\iidct+\isumo-2}\\
    & + \Afac[\LamPhi] \underbrace{\PolyP[\lambdadt]{0}{1}}_{=1} \eep{\iidct - 1} + \left(\Afac[\LamPhi] \underbrace{\PolyQ[\lambdadt]{0}{1}}_{=0} + 1\right)\ee{\iidct - 1}\\
    \le & \sum_{\isumo=1}^{\then-\iidct + 1}\glb\PolyP[\lambdadt]{1}{\isumo}\eep{\iidct +\isumo - 2} + \PolyQ[\lambdadt]{1}{\isumo}\ee{\iidct + \isumo - 2}\grb\\
    = & \Bound{1}{\iidct - 1}.
\end{align*}
}
Thus, we proved that $\EEP{\iidct - 1}$ and $\EE{\iidct - 1}$ are bounded by $\Bound{0}{\iidct - 1}$ and $\Bound{1}{\iidct - 1}$ respectively. Applying the relation $\mathbb{E}^h [\cdot]\le\sqrt{\mathbb{E}^h [\cdot^2]}$, we obtain \eqref{e:regerr_general_bound} and prove Theorem~\ref{t:induction}.
 
  \bibliographystyle{chicago}\bibliography{ref}
\end{document}